\documentclass[12pt]{article}
\usepackage[title]{appendix}
\usepackage{amsmath}
\usepackage{breqn}
\usepackage{color, colortbl}
\usepackage{amsfonts}
\usepackage{amssymb}
\usepackage{graphicx}
\usepackage[round]{natbib}
\usepackage{float}
\usepackage{multirow}
\usepackage{tikz}
\usepackage{hyperref}
\usepackage{algorithm}
\usepackage{mathrsfs}

\usepackage{subcaption}
\usepackage[noend]{algpseudocode}

\usepackage{rotating}
\usepackage{setspace} 
\usepackage{multicol}
\usepackage[margin=1in]{geometry}

\definecolor{lightGray}{gray}{0.97}
\definecolor{medGray}{gray}{0.87}
\definecolor{darkGray}{gray}{0.77}

\usepackage{tabularx}
\usepackage{makecell}
\usepackage{array}
\usepackage{epsfig}
\usepackage{svg}

\usepackage{booktabs}

\title{Parsimonious Ultrametric Manly Mixture Models}
\author{Alexa A. Sochaniwsky  and Paul D. McNicholas}
\date{\small Department of Mathematics \& Statistics, McMaster University,\\ Ontario, Canada}

\begin{document}

\maketitle 

\begin{abstract}
A family of parsimonious ultrametric mixture models with the Manly transformation is developed for clustering high-dimensional data where the clusters may be asymmetric. While advances in Gaussian mixture modeling sufficiently handle high-dimensional data, they often struggle with the common presence of cluster skewness. To address this, we incorporate the extended ultrametric covariance structure and the Manly transformation, resulting in the parsimonious ultrametric Manly mixture model family. The ultrametric covariance structure reduces the number of free parameters while identifying latent groups of variables within a nested hierarchy. This phenomenon enables the visualization of hierarchical relationships within clusters, improving cluster interpretability. Additionally, as with many classes of mixture models, model selection remains a fundamental challenge; to this end, a two-step model selection procedure is proposed herein. Through simulation studies and real data analyses, we demonstrate improved model selection via the proposed two-step method, as well as the effective clustering performance.
\end{abstract}

\section{Introduction}
Finite mixture models are a popular clustering method due to their ability to model sub-populations. In the parametric paradigm, we assume that finite $G$ sub-populations exist and the observations belonging to each group can be modeled by a probabilistic distribution. A $p$-dimensional random vector $\mathbf{X}$ is from a $G$-component mixture model if the density can be written 
    \begin{equation*}
f(\mathbf{x}\mid\boldsymbol{\vartheta}) = \sum_{g = 1}^{G}\pi_g f_g(\mathbf{x}\mid\boldsymbol{\theta}_g),
\end{equation*}
where $\pi_g > 0$ is the $g$th mixing proportion with $\sum_{g = 1}^{G} \pi_g = 1$, $f_g(\mathbf{x}\mid\boldsymbol{\theta}_g)$ is the $g$th component density, and $\boldsymbol{\vartheta} = ( \pi_1,..., \pi_G,\boldsymbol{\theta}_1,...,\boldsymbol{\theta}_G)$ is a vector of the parameters. Gaussian mixture models have been and continue to be prominent in the literature. The $g$th Gaussian component density is denoted as $\phi(\mathbf{x}|\boldsymbol{\mu}_g, \boldsymbol{\Sigma}_g)$ where $\boldsymbol{\mu}_g$ is the mean vector and $\boldsymbol{\Sigma}_g$ is the covariance matrix. The standard for estimating these models is via the expectation-maximization (EM) algorithm \citep{dempster77} as demonstrated in multiple monographs; see, e.g., \citet{titterington85}, \citet{mclachlan00}, and \citet{mcnicholas2016}. 

A challenge for standard parametric mixture models is the number of free parameters that need to be estimated, particularly when fitting high-dimensional data. Gaussian mixture models, for example, estimate a total of $(G-1) + Gp + Gp(p+1)/2$ free parameters, with $G-1$ being from the mixing proportions, $Gp$ from the means, and $Gp(p+1)/2$ from the covariance matrices. There have been a variety of approaches to decrease this number including implementing matrix decompositions and data reduction techniques. \citet{banfield1993} and \citet{celeux95} use an eigen-decomposition for the component covariances such that $\boldsymbol{\Sigma}_g = \lambda_g \mathbf{D}_g \mathbf{A}_g \mathbf{D}_g^{\top}$, where $\lambda_g = |\boldsymbol{\Sigma}_g|^{1/p}$, $\mathbf{D}_g$ is a matrix of eigenvectors, and $\mathbf{A}_g$ is a diagonal matrix of eigenvalues with $|\mathbf{A}_g|=1$. Constraining $\boldsymbol{\Sigma}_g$ across parameters and components results in the Gaussian parsimonious clustering models \citep[GPCMs;][]{celeux95}. \citet{mcnicholas2008,mcnicholas10d} propose a class of models that includes mixtures of factor analyzers \citep{ghahramani97} and mixtures of probabilistic principal component analyzers \citep{tipping99} such that the component covariance is of the form $\boldsymbol{\Sigma}_g = \boldsymbol{\Lambda}_g \boldsymbol{\Lambda}_g^{\top} + \boldsymbol{\Psi}_g$, where $\boldsymbol{\Lambda}_g$ is a factor loading matrix and $\boldsymbol{\Psi}_g$ is a diagonal covariance matrix of error terms. Imposing constraints on $\boldsymbol{\Lambda}_g$ and $\boldsymbol{\Psi}_g$ results in the parsimonious Gaussian mixture model (PGMM) family. Most recently, \citet{cavicchia2024} constrain the parameters in the extended ultrametric covariance structure introduced by \cite{cavicchia2022} introducing the parsimonious ultrametric Gaussian mixture model (PUGMM) family. These models not only decrease the number of free parameters, but each model is associated with a hierarchy of latent groups. Parameter estimation is carried out via a grouped coordinate ascent algorithm \citep{zangwill69} with the Hathaway log-likelihood function \citep{hathaway1986}.

Despite the popularity of a Gaussian component density, it is not suitable for all data. For example, Gaussian mixture models tend to overestimate the number of clusters when clusters are asymmetric or have heavier tails \citep[see][for examples]{franczak2014, mcnicholas2016}. There have been advances in model-based clustering to address these shortcomings, including mixtures with flexible component densities and the use of transformations to near-normality. Mixtures of flexible distributions include the normal-inverse Gaussian \citep{karlis09,ohagan16}, skew-$t$ \citep{lin2009,vrbik14,murray2014}, shifted asymmetric Laplace \citep{franczak2014}, generalized hyperbolic \citep{browne2015}, power exponential  \citep{dang15}, and skewed power exponential \citep{dang23}. For dimension reduction, mixtures of factor analyzers for generalized hyperbolic distributions \citep{tortora2016}, the skew-$t$ distribution \citep{murray2014}, the variance-gamma distribution \citep{mcnicholas2017}, and the hidden truncation hyperbolic distribution \citep{murray20} have been developed. Mixtures of transformations include the Box-Cox \citep{lo2012}, Manly \citep{zhu2018}, and Yeo-Johnson \citep{cai2024}. Despite these advances, options for parsimonious modeling of asymmetric data remain limited.

The work of \citet{cavicchia2024} is extended herein to asymmetric clusters by proposing the parsimonious ultrametric Manly mixture models (PUMMMs).

\section{Background}
\label{Sec2}
\subsection{Extended Ultrametric Covariance Structure} \label{Sec21}
To reduce the number of free covariance parameters, \citet{cavicchia2022} introduce the extended ultrametric covariance structure (EUCovS), merging $p$ variables into $m$ latent groups, arranged in a nested hierarchy. The EUCovS is defined as 
\begin{equation}
\label{EUCovS}
    \boldsymbol{\Sigma}_g = \mathbf{V}_g \left(\boldsymbol{\Sigma}_{W_g} + \boldsymbol{\Sigma}_{B_g} \right) \mathbf{V}_g^{\top} + \text{diag}\left\{\mathbf{V}_g \left( \boldsymbol{\Sigma}_{V_g} - \boldsymbol{\Sigma}_{W_g} \right) \mathbf{V}_g^{\top} \right\},
\end{equation}
where $\mathbf{V}_g = \left[v_{jq} : j = 1, \hdots, p, q= 1, \hdots, m\right]$ is a binary and row-stochastic group variable membership matrix with $m \leq p$ groups, $\boldsymbol{\Sigma}_{V_g} = \left[_V \sigma_{qq(g)} : q = 1,\hdots,m\right]$ is a diagonal group variance matrix, $\boldsymbol{\Sigma}_{W_g} = \left[_W \sigma_{qq(g)} : q = 1,\hdots,m\right]$ is a diagonal within-group covariance matrix, and  $\boldsymbol{\Sigma}_{B_g} = \left[_B \sigma_{qh(g)} : q,h = 1,\hdots,m\right]$ is a symmetric between-group covariance matrix. The matrices that comprise $\boldsymbol{\Sigma}_g$ must comply with the following constraints to guarantee ultrametricity: 
\begin{enumerate}
    \item[(i)] $_B \sigma_{qh(g)} \geq \min \{ _B \sigma_{qs(g)}, _B \sigma_{hs(g)} \}$, $q,h,s = 1,\hdots, m, s \neq h \neq q$,
    \item[(ii)] $\min \{_W\sigma_{qq(g)}, q = 1,\hdots, m\} \geq  \max \{_B\sigma_{qh(g)}, q,h = 1,\hdots, m, h \neq q\}$,
    \item[(iii)] $_V \sigma_{qq(g)} > |_W \sigma_{qq(g)} |$, $q = 1,\hdots,m$.
\end{enumerate}
When all covariance entries are nonnegative, $\boldsymbol{\Sigma}_g$ is positive semidefinite. When entries are not nonnegative, the positive-semidefinite requirement of the covariance is guaranteed using the polar decomposition \citep{higham1986} described in \citet{cavicchia2024}. 

The ultrametric covariance is associated with a hierarchical structure where variables merged earlier in the hierarchy have a stronger relationship than variables merged later. Each group is characterized by its group variance, within-group covariance, and between-group covariances. Specifically, $\boldsymbol{\Sigma}_{V_g}$ defines the initial positions in the hierarchy and is interpreted as the group’s initial average level of consistency. The $2m - 1$ aggregation levels are determined by $\boldsymbol{\Sigma}_{W_g}$ and $\boldsymbol{\Sigma}_{B_g}$ such that $\boldsymbol{\Sigma}_{W_g}$ defines the levels at which variables are aggregated, while the remaining $m - 1$ levels are determined by $\boldsymbol{\Sigma}_{B_g}$. The aggregation levels represent a measure of concordance among groups, with groups progressing from most to least concordant as they are combined. This parameterization assumes that hierarchical relationships between the variables exist as ultrametricity is guaranteed, and that the number of variable groups, $m$, is equal across clusters.

\citet{cavicchia2024}  propose the PUGMM family, which arises by constraining the parameters in the EUCovS. The PUGMM family consists of thirteen models, which can be divided into two groups: the unique and equal models (EUUU, EUUE, EUEE, EEEU, and EEEE); and the isotropic and free models (EEEF, EEFF, EFFF, FIII, FIIF, FIFF, FFFI, and FFFF). The nomenclature of the models as they pertain to $\boldsymbol{\Sigma}_{V_g}$, $\boldsymbol{\Sigma}_{W_g}$ and $\boldsymbol{\Sigma}_{B_g}$ are defined as: equal (E) corresponds to the parameter being equal across components, unique (U) corresponds to the parameter being equal within and across components, isotropic (I) corresponds to the elements being equal within a parameter, and free (F) corresponds to the parameter varying across components. Only E and F are relevant to the variable-group membership matrix $\mathbf{V}_g$. The models are detailed in Table \ref{table:pummm}.

\begin{sidewaystable}[p]
    \centering
    \caption{The model case, nomenclature, covariance structure, and number of covariance parameters of the PUGMMs or the PUMMMs.}
    \label{table:pummm}
    \scalebox{1}{\small{
		\begin{tabular*}{1.0\textwidth}{@{\extracolsep{\fill}}lcccll} 
		  \noalign{\smallskip}\hline
           Model & \multicolumn{3}{c}{Nomenclature} & Covariance Structure & $\#$ of Covariance\\ 
\cline{2-4}
		                 & $\mathbf{\Sigma}_{V_g}$ &$\mathbf{\Sigma}_{W_g}$ & $\mathbf{\Sigma}_{B_g}$ &  & Parameters\\
\hline 
$\boldsymbol{V}$ Equal \\
[0.7ex]
EUUU & Unique & Unique & Unique & $\boldsymbol{V}(\sigma_W \mathbf{I}_m + \sigma_B (\mathbf{1}_m \mathbf{1}_m^{\top} - \mathbf{I}_m))\boldsymbol{V}^{\top} + \text{diag}(\boldsymbol{V}(\sigma_V - \sigma_W) \mathbf{I}_m\boldsymbol{V}^{\top})$ & $p + 3$ \\
EUUE & Unique & Unique & Equal & $\boldsymbol{V}(\sigma_W \mathbf{I}_m + \boldsymbol{\Sigma}_B)\boldsymbol{V}^{\top} + \text{diag}(\boldsymbol{V}(\sigma_V - \sigma_W) \mathbf{I}_m\boldsymbol{V}^{\top})$ & $p + m + 1$ \\
EUEE & Unique & Equal & Equal & $\boldsymbol{V}(\boldsymbol{\Sigma}_W + \boldsymbol{\Sigma}_B)\boldsymbol{V}^{\top} +\text{diag}(\boldsymbol{V}(\sigma_V\mathbf{I}_m - \boldsymbol{\Sigma}_W) \boldsymbol{V}^{\top})$ & $p + 2m$ \\
EEEU & Equal & Equal & Unique & $\boldsymbol{V}(\boldsymbol{\Sigma}_W  + \sigma_B (\mathbf{1}_m \mathbf{1}_m^{\top} - \mathbf{I}_m))\boldsymbol{V}^{\top} + \text{diag}(\boldsymbol{V}(\boldsymbol{\Sigma}_V - \boldsymbol{\Sigma}_W ) \boldsymbol{V}^{\top})$ & $p + 2m + 1$ \\
EEEE & Equal & Equal & Equal & $\boldsymbol{V}(\boldsymbol{\Sigma}_W + \boldsymbol{\Sigma}_B)\boldsymbol{V}^{\top} +\text{diag}(\boldsymbol{V}(\boldsymbol{\Sigma}_V - \boldsymbol{\Sigma}_W ) \boldsymbol{V}^{\top})$ &$p + 3m - 1$ \\
EEEF & Equal & Equal & Free & $\boldsymbol{V}(\boldsymbol{\Sigma}_W + \boldsymbol{\Sigma}_{B_g})\boldsymbol{V}^{\top} +\text{diag}(\boldsymbol{V}(\boldsymbol{\Sigma}_V - \boldsymbol{\Sigma}_W ) \boldsymbol{V}^{\top})$  &$p + 2m + G(m-1)$ \\
EEFF & Equal & Free & Free & $\boldsymbol{V}(\boldsymbol{\Sigma}_{W_g} + \boldsymbol{\Sigma}_{B_g})\boldsymbol{V}^{\top} +\text{diag}(\boldsymbol{V}(\boldsymbol{\Sigma}_V - \boldsymbol{\Sigma}_{W_g}) \boldsymbol{V}^{\top})$ & $p + m + G(2m-1)$ \\
EFFF  & Free & Free & Free & $\boldsymbol{V}(\boldsymbol{\Sigma}_{W_g} + \boldsymbol{\Sigma}_{B_g})\boldsymbol{V}^{\top} +\text{diag}(\boldsymbol{V}(\boldsymbol{\Sigma}_{V_g} - \boldsymbol{\Sigma}_{W_g}) \boldsymbol{V}^{\top})$ & $p + G(3m-1)$ \\
[0.7ex]
$\boldsymbol{V}$ Free \\
[0.7ex]
FIII  & Isotropic & Isotropic & Isotropic & $\boldsymbol{V}_g(\sigma_{W_g} \mathbf{I}_m + \sigma_{B_g} (\mathbf{1}_m \mathbf{1}_m^{\top} - \mathbf{I}_m))\boldsymbol{V}_g^{\top} + \text{diag}(\boldsymbol{V}_g(\sigma_{V_g} - \sigma_{W_g}) \mathbf{I}_m\boldsymbol{V}_g^{\top})$ & $G (p + 3)$ \\
FIIF  & Isotropic & Isotropic & Free & $\boldsymbol{V}_g(\sigma_{W_g} \mathbf{I}_m + \boldsymbol{\Sigma}_{B_g})\boldsymbol{V}_g^{\top} + \text{diag}(\boldsymbol{V}_g(\sigma_{V_g} - \sigma_{W_g}) \mathbf{I}_m\boldsymbol{V}_g^{\top})$ & $G (p + m + 1)$ \\
FIFF  & Isotropic & Free & Free & $\boldsymbol{V}_g(\boldsymbol{\Sigma}_{W_g} + \boldsymbol{\Sigma}_{B_g})\boldsymbol{V}_g^{\top} +\text{diag}(\boldsymbol{V}_g(\sigma_{V_g}\mathbf{I}_m - \boldsymbol{\Sigma}_{W_g}) \boldsymbol{V}_g^{\top})$ & $G (p + 2m)$ \\
FFFI & Free & Free & Isotropic & $\boldsymbol{V}_g (\boldsymbol{\Sigma}_{W_g}  + \sigma_{B_g} (\mathbf{1}_m \mathbf{1}_m^{\top} - \mathbf{I}_m))\boldsymbol{V}_g^{\top} + \text{diag}(\boldsymbol{V}_g(\boldsymbol{\Sigma}_{V_g} - \boldsymbol{\Sigma}_{W_g} ) \boldsymbol{V}_g^{\top})$ & $G (p + 2m + 1)$ \\
FFFF  & Free & Free & Free & $\boldsymbol{V}_g (\boldsymbol{\Sigma}_{W_g} + \boldsymbol{\Sigma}_{B_g})\boldsymbol{V}_g^{\top} +\text{diag}(\boldsymbol{V}_g(\boldsymbol{\Sigma}_V - \boldsymbol{\Sigma}_{W_g}) \boldsymbol{V}_g^{\top})$ & $G (p + 3m - 1)$ \\
\hline
\end{tabular*}}}
\end{sidewaystable}

\subsection{Manly Transformation}
One transformation to normality is the exponential transformation introduced by \citet{manly1976}. For a scalar variable $x$, the Manly transformation is defined as
\[ y = \begin{cases} 
   \frac{e^{\lambda x} - 1}{\lambda}, & \lambda \neq 0 \\
      x, &\lambda = 0,
   \end{cases}
\]
where $\lambda$ is the transformation parameter and $y$ is the transformed variable. As a result, $\lambda$ restricts the support of $y$: for $\lambda > 0$, $y$ has range $-1/\lambda$ to $\infty$, and for $\lambda < 0$, $y$ has range $-\infty$ to $-1/\lambda$, a trade-off for near-normality.

Prior to the development of the Manly transformation, the Box-Cox transformation \citep{box1964} was, and remains, one of the most widely used transformations; however, it can only be applied to $x\in\mathbb{R}^+$ and is primarily effective on right-skewed data. In contrast, the Manly transformation can be applied to $x\in\mathbb{R}$ for both right- and left-skewed data. More recently, \citet{yeo2000} proposed the Yeo–Johnson transformation, which, like the Manly transformation, is defined for $x\in\mathbb{R}$. However, following the discussion in \citet{zhu2018}, we implement the Manly transformation due to its mathematical convenience.

\subsection{Manly Mixture Models}
In \citet{zhu2018}, the multivariate Manly transformation is adapted for mixture models by assuming that, for the $g$th component, there exists a transformation vector $\boldsymbol{\lambda}_g  = (\lambda_{g1}, \hdots , \lambda_{gp})$ that results in
\begin{equation*}
    \mathbf{Y}_g = \left( \frac{e^{\lambda_{g1}\mathbf{x}_1} - 1}{\lambda_{g1}}, \hdots,  \frac{e^{\lambda_{gp}\mathbf{x}_p} - 1}{\lambda_{gp}} \right) \sim N_p(\boldsymbol{\mu}_g, \boldsymbol{\Sigma}_g),
\end{equation*}
where $\mathbf{Y}_g$ denotes the transformed data corresponding to $\boldsymbol{\lambda}_g$.  The original data, $\mathbf{x}$, is then of a Manly mixture model if its density can be written as 
\begin{equation*}
    f(\mathbf{x} | \boldsymbol{\vartheta}) = \sum_{g = 1}^G \pi_g \phi(\mathcal{M}(\mathbf{x};\boldsymbol{\lambda}_g) \mid \boldsymbol{\mu}_g, \boldsymbol{\Sigma}_g) \exp(\boldsymbol{\lambda}_g^{\top}\mathbf{x}),
\end{equation*} 
where $\exp(\boldsymbol{\lambda}_g^{\top}\mathbf{x})$ is the Jacobian as a result of the transformation, and $\mathcal{M}(\mathbf{X};\boldsymbol{\lambda}_g)  \equiv \mathbf{Y}_g$ denotes the Manly operator.

\section{Methodology}
\label{Sec3}

\subsection{Overview}
\label{Manly_}

The PUMMM family is introduced herein, in analogy to the PUGMM family of \citet{cavicchia2024}, with a novel two-step model selection method. By implementing the EUCovS and the Manly transformation, the proposed models introduce parsimony and identify hierarchical relationships between groups of variables while also modeling skewness. Constraining the parameters in the EUCovS results in a total of thirteen covariance cases. As with the PUGMMs, the PUMMMs contain unique and equal cases, EUUU, EUUE, EUEE, EEEU, EEEE, and the isotropic and free cases, EEEF, EEFF, EFFF, FIII, FIIF, FIFF, FFFI, FFFF. All models are as defined in Table~\ref{table:pummm}.

\subsection{Parameter Estimation}
A $G$-component parsimonious ultrametric Manly mixture model has density
\begin{equation*}
    f(\mathbf{x} | \boldsymbol{\vartheta}) = \sum_{g = 1}^G \pi_g \phi( \mathcal{M}(\mathbf{x};\boldsymbol{\lambda}_g) \mid \boldsymbol{\mu}_g, \mathbf{V}_g, \boldsymbol{\Sigma}_{V_g}, \boldsymbol{\Sigma}_{W_g},\boldsymbol{\Sigma}_{B_g})\exp(\boldsymbol{\lambda}_g^{\top}\mathbf{x}),
\end{equation*}
where $\mathbf{V}_g$, $\boldsymbol{\Sigma}_{V_g}$, $\boldsymbol{\Sigma}_{W_g}$, and $\boldsymbol{\Sigma}_{B_g}$ are all of a specified ultrametric covariance structure.
Parameter estimation is carried out using a grouped coordinate ascent algorithm \citep{zangwill69}. \citet{hathaway1986} showed that, for Gaussian mixture models, this algorithm is equivalent to the EM algorithm, the most common approach for parametric finite mixture models. The Hathaway log-likelihood comprises a complete-data log-likelihood and an entropy term where the posterior probabilities are considered model parameters. As a result of the entropy term, maximizing the Hathaway log-likelihood with respect to the posterior probabilities is equivalent to the E-step in the EM algorithm. Maximizing the Hathaway log-likelihood with respect to the remaining model parameters is equivalent to the parameter updates in the M-step because the entropy term is constant with respect to said model parameters. 

This algorithmic equivalence is preserved in the proposed family. The Jacobian as a result of the Manly transformation does not interfere with the posterior probability derivations of \citet{hathaway1986}, and the model parameter updates, including the transformation parameter, result from coordinate ascent on the Hathaway log-likelihood. The Hathaway log-likelihood for the PUMMMs is given by
\begin{dmath}
    \ell (\boldsymbol{Z}, \boldsymbol{\vartheta}) = \sum_{i=1}^n \sum_{g=1}^G z_{ig} \left[ \log \pi_g + \log \phi(\mathcal{M}(\mathbf{x}_i;\boldsymbol{\lambda}_g) | \boldsymbol{\mu}_g, \boldsymbol{\Sigma}_g) + \boldsymbol{\lambda}_g^{\top}\mathbf{x}_i \right] - \sum_{i=1}^n \sum_{g=1}^G z_{ig} \log z_{ig},
    \label{CDLL}
\end{dmath}
where $z_{ig} \in [0,1]$ denote the posterior probabilities  subject to $\sum_{g=1}^G z_{ig} = 1$, and $\sum_{i=1}^n \sum_{g=1}^G z_{ig} = n$. 

As described, parameter estimates are obtained by maximizing $\ell (\boldsymbol{Z}, \boldsymbol{\vartheta})$ with respect to $\boldsymbol{Z}$ and, then with respect to each parameter in $\boldsymbol{\vartheta}$ while holding the other parameters fixed. For simplicity, we denote $\ell (\boldsymbol{Z}, \boldsymbol{\vartheta})$ as $\ell$ hereafter.
Starting with the posterior probabilities, for $i = 1, \hdots, n$ and $g = 1, \hdots, G$, the update for $z_{ig}$ is
\begin{equation*}
    \hat{z}_{ig} = \frac{\hat{\pi}_g \phi( \mathcal{M}(\mathbf{x}_i;\hat{\boldsymbol{\lambda}}_g) | \hat{\boldsymbol{\mu}}_g, \hat{\boldsymbol{\Sigma}}_g) \exp(\hat{\boldsymbol{\lambda}}_g^{\top}\mathbf{x}_i)}{\sum_{h = 1}^G \hat{\pi}_h \phi( \mathcal{M}(\mathbf{x}_i;\hat{\boldsymbol{\lambda}}_h) | \hat{\boldsymbol{\mu}}_h, \hat{\boldsymbol{\Sigma}}_h) \exp(\hat{\boldsymbol{\lambda}}_h^{\top}\mathbf{x}_i)},
\end{equation*}
where the covariance structure is case dependent (see Table~\ref{table:pummm}).
The update for the transformation parameter $\boldsymbol{\lambda}_g$ is obtained by maximizing the expression
\begin{equation*}
    \sum_{i = 1}^n \hat{z}_{ig} \left[\phi(\mathcal{M}(\mathbf{x}_i;\boldsymbol{\lambda}_g) ; \boldsymbol{\mu}_g, \boldsymbol{\Sigma}_g) + \boldsymbol{\lambda}_g^{\top}\mathbf{x}_i \right]
\end{equation*}
with respect to $\boldsymbol{\lambda}_g$. 
Various optimization methods can be used to estimate $\boldsymbol{\lambda}_g$ but the Nelder-Mead method \citep{nelder1965} is implemented herein. Updates for the mixing proportions $\hat{\pi}_g$ and means $\hat{\boldsymbol{\mu}}_g$ are given by
\begin{equation*}
     \hat{\pi}_g  = \frac{n_g}{n} \hspace{5mm} \text{ and } \hspace{5mm}
     \hat{\boldsymbol{\mu}}_g  = \frac{1}{n_g} \sum_{i =1}^n \hat{z}_{ig} \hspace{1mm} \mathcal{M}(\mathbf{x}_i;\hat{\boldsymbol{\lambda}}_g),
\end{equation*}
respectively, where $n_g = \sum_{i = 1}^n \hat{z}_{ig}$. 

The $\hat{\boldsymbol{\Sigma}}_g$ update is dependent on the PUMMM being estimated. Based on the specified PUMMM, each covariance structure has its own set of updates for $\mathbf{V}_g$, $\boldsymbol{\Sigma}_{V_g}$, $\boldsymbol{\Sigma}_{W_g}$, and $\boldsymbol{\Sigma}_{B_g}$. The update for the variable-group membership matrix is as follows.
For the models beginning with~E, the variable-group membership parameter is constrained such that $\mathbf{V}_g = \mathbf{V}$ and the update for each row across components $\mathbf{v}_j$, $j = 1, \hdots, p$, is 
\begin{equation*}
    \begin{cases} 
      \hat{v}_{jq} = 1 & \text{if } q = \underset{q' = 1, \ldots, m}{\text{arg }\text{max }} \ell_{-\mathbf{V}},\\
      \hat{v}_{jq} = 0 & \text{otherwise,}
    \end{cases}
\end{equation*}
where $\ell_{-\mathbf{V}} = \ell(\hat{\boldsymbol{Z}}, \hat{\boldsymbol{\vartheta}}_{-\mathbf{V}}, [\hat{\mathbf{v}}_1, \hdots , \hat{\mathbf{v}}_j = \mathbf{i}_{q'}, \hdots \hat{\mathbf{v}}_p]^{\top})$ such that $\hat{\boldsymbol{\vartheta}}_{-\mathbf{V}}$
contains all the model parameters except for $\mathbf{V}$, and $\mathbf{i}_{q'}$ is the $q'$th row of an $m$ order identity matrix. For the models beginning with F, $\mathbf{V}_g$ is free to vary across components and the update for each row across components $\mathbf{v}_{j (g)}$, $j = 1, \hdots, p$, $g = 1, \hdots, G$, is 
\begin{equation*}
    \begin{cases} 
      \hat{v}_{jq (g)} = 1 & \text{if } q = \underset{q' = 1, \hdots, m}{\text{arg } \text{max }} \ell_{-\mathbf{V}_g},\\
      \hat{v}_{jq (g)} = 0 & \text{otherwise,}
    \end{cases}
\end{equation*}
where $\ell_{-\mathbf{V}_g} = \ell(\hat{\boldsymbol{Z}},\hat{\boldsymbol{\vartheta}}_{-\mathbf{V}_g}, [\hat{\mathbf{v}}_{1(g)}, \hdots , \hat{\mathbf{v}}_{j(g)} = \mathbf{i}_{q'(g)}, \hdots \hat{\mathbf{v}}_{p(g)}]^{\top})$ such that $\hat{\boldsymbol{\vartheta}}_{-\mathbf{V}_g}$
contains all the model parameters except $\mathbf{V}_g$.

As aforementioned, the PUMMM family can be divided into two groups: the unique and equal models; and the isotropic and free models.
For the unqiue and equal models, $\boldsymbol{\Sigma}_g = \boldsymbol{\Sigma}$ and
the updates for $\boldsymbol{\Sigma}_{V_g}$, $\boldsymbol{\Sigma}_{W_g}$, and $\boldsymbol{\Sigma}_{B_g}$ arise from maximizing
\begin{equation*}
    \ell = -\frac{n}{2} (\log | \boldsymbol{\Sigma}|+ \text{tr}\{\boldsymbol{\Sigma}^{-1} \bar{\mathbf{S}}\}),
\end{equation*}
where $\bar{\mathbf{S}} = \sum_{g = 1}^G \hat{\pi}_g \mathbf{S}_g$ and 
\begin{equation}\label{eqn:sampcov}
    \mathbf{S}_g = \frac{1}{n_g} \sum_{i = 1}^n \hat{z}_{ig} (\mathcal{M}(\mathbf{x}_i;\hat{\boldsymbol{\lambda}}_g) - \hat{\boldsymbol{\mu}}_g)(\mathcal{M}(\mathbf{x}_i;\hat{\boldsymbol{\lambda}}_g) - \hat{\boldsymbol{\mu}}_g)^{\top}. 
\end{equation}
For the unique and free models, the updates for $\boldsymbol{\Sigma}_{V_g}$, $\boldsymbol{\Sigma}_{W_g}$, and $\boldsymbol{\Sigma}_{B_g}$ arise from maximizing 
\begin{equation*}
    \ell = -\frac{1}{2} \left(\sum_{g = 1}^G n_g \log | \boldsymbol{\Sigma}_g|+ \sum_{g = 1}^G n_g \text{tr}\{\boldsymbol{\Sigma}_g^{-1} \mathbf{S}_g\}\right).
\end{equation*}
The parameter updates for $\boldsymbol{\Sigma}_{V_g}$, $\boldsymbol{\Sigma}_{W_g}$, and $\boldsymbol{\Sigma}_{B_g}$ are analogous to those presented in \citet{cavicchia2024} with the $\mathbf{S}_g$ as defined in \eqref{eqn:sampcov}; accordingly, they are not reproduced herein.

\subsection{Model Selection}
In model-based clustering literature, whether for Gaussian or skewed distributions, the Bayesian information criterion \citep[BIC;][]{schwarz1978} is the most commonly used model selection criterion. This selection is also reflected in various {\sf R} \citep{R} packages for mixture models. Specifically, the BIC is the default model selection criterion in both \texttt{mixture} \citep{mixture} and \texttt{mclust} \citep{mclust}, which implement the GPCM family, in \texttt{pgmm} \citep{pgmm}, which implements the PGMM family, and in \texttt{PUGMM} \citep{PUGMM}, which implements the PUGMM family. The BIC can be written as
\begin{equation*}
    \text{BIC} = 2 \ell(\hat{\boldsymbol{Z}}, \hat{\boldsymbol{\vartheta}})- \rho \log n,
\end{equation*}
where $\ell (\hat{\boldsymbol{Z}}, \hat{\boldsymbol{\vartheta}})$ is the maximized log-likelihood, $\rho$ is the number of free parameters, and $n$ is the number of observations. The model with the maximum BIC is selected.

While the BIC has been shown to be an effective criterion \citep{fraley02}, we found it to be inconsistent in its ability to select the correct model, i.e., the number of components $G$, the number of groups of variables $m$, and the covariance case, denoted as ($G$, $m$, case). When an exhaustive model search is done, the BIC often overestimates $m$ and incorrectly selects the covariance case. To demonstrate the ineffectiveness of the BIC in selecting the correct model, we use Occam's window \citep{madigan1994}. With Occam's window, one obtains a subset of ``good'' models, where models that are considered to have a much worse fit than the best model are discarded. 
\citet{wei2015} have that Occam's window is equivalent to 
\begin{equation*}
    \{M_i : \max\{\text{BIC}_I\} - \text{BIC}_i  \leq 2 \log c\},
\end{equation*}
where $M_i$ is a given model, $\text{BIC}_i$ is the BIC value corresponding to model $M_i$, $\text{BIC}_I$ is the BIC corresponding to the model with the maximum BIC, and $c$ is a positive constant. We follow \citet{madigan1994} by using $c = 20$ herein. Using simulated data from Simulation 1, we consider the number of times the correct model was captured within Occam's window, and the number of times the correct model was selected by the BIC (Table~\ref{table:occams}). In Simulation 1, the data are generated from a three-component Manly mixture with an ultrametric covariance structure and the transformation parameter being one of the lambda settings described therein, denoted by $\boldsymbol{\lambda}^{(k)}$ for $k = 1,2,3$.
\begin{table}[!htb]
\centering
\caption{Percentage of times the correct model is selected in Occam's window and by BIC.}
\small
\scalebox{1}{
\begin{tabular*}{0.75\textwidth}{@{\extracolsep{\fill}} cccccc}
\toprule
 &  & 
\multicolumn{2}{c}{Occam's Window} & 
\multicolumn{2}{c}{BIC} \\
\cline{3-4}\cline{5-6}
$\boldsymbol{\lambda}^{(k)}$ & Model & $n = 250$ & $n = 500$ & $n = 250$ & $n = 500$ \\
\midrule
\multirow{4}{*}{$\boldsymbol{\lambda}^{(1)}$}
& EUUU & \cellcolor{gray!10}92.00\% & \cellcolor{gray!20}94.00\%\% & \cellcolor{gray!10}42.00\% & \cellcolor{gray!20}46.67\% \\
& EEEE & \cellcolor{gray!10}17.33\% & \cellcolor{gray!20} 44.00\% & \cellcolor{gray!10} 4.00\% & \cellcolor{gray!20} 18.67\% \\
& FIII & \cellcolor{gray!10}74.67\% & \cellcolor{gray!20} 74.67\% & \cellcolor{gray!10} 63.33\% & \cellcolor{gray!20} 69.33\% \\
& FFFF & \cellcolor{gray!10}1.33 \% & \cellcolor{gray!20} 28.67\% & \cellcolor{gray!10} 0.00\%  & \cellcolor{gray!20} 20.67\% \\
\midrule
\multirow{4}{*}{$\boldsymbol{\lambda}^{(2)}$}
& EUUU & \cellcolor{gray!10} 94.00\% & \cellcolor{gray!20} 97.33\% & \cellcolor{gray!10} 44.67\% & \cellcolor{gray!20} 50.67\% \\
& EEEE & \cellcolor{gray!10} 20.00\% & \cellcolor{gray!20} 46.00\% & \cellcolor{gray!10} 6.00\%  & \cellcolor{gray!20} 24.67\% \\
& FIII & \cellcolor{gray!10} 73.33\% & \cellcolor{gray!20} 76.67\% & \cellcolor{gray!10} 62.00\% & \cellcolor{gray!20} 70.67\% \\
& FFFF & \cellcolor{gray!10} 2.67\%  & \cellcolor{gray!20} 34.00\% & \cellcolor{gray!10} 0.67\%  & \cellcolor{gray!20} 27.33\% \\

\midrule
\multirow{4}{*}{$\boldsymbol{\lambda}^{(3)}$}
& EUUU & \cellcolor{gray!10}95.33\% & \cellcolor{gray!20} 97.33\% & \cellcolor{gray!10} 43.33\% & \cellcolor{gray!20} 48.67\% \\
& EEEE & \cellcolor{gray!10} 16.67\% & \cellcolor{gray!20} 46.67\% & \cellcolor{gray!10} 8.67\%  & \cellcolor{gray!20} 22.00\% \\
& FIII & \cellcolor{gray!10} 80.67\% & \cellcolor{gray!20} 76.00\% & \cellcolor{gray!10} 67.33\% & \cellcolor{gray!20} 72.00\% \\
& FFFF & \cellcolor{gray!10} 4.67\%  & \cellcolor{gray!20} 31.33\% & \cellcolor{gray!10} 1.33\%  & \cellcolor{gray!20} 26.67\% \\

\bottomrule
\end{tabular*}}
\label{table:occams}
\end{table}

From Table~\ref{table:occams}, we can see that the results are inconsistent, varying across $n$ and model. Specifically, for the EUUU case, the correct model ($G$, $m$, case) is within Occam's window about 95\% of the time but the BIC only selects the correct model about 45\% of the time.

Seeing as Occam's window can include the correct model in the subset of ``good'' models a substantial amount of the time, these results justify a variation of the BIC model selection approach. Other criteria were investigated, namely, the integrated completed likelihood \citep[ICL;][]{biernacki2000} and the Akaike information criterion \citep[AIC;][]{akaike2011} but similar inconsistencies occurred. Thus, we propose a two-step model selection method. Let $G^*$ denote the range of values for $G$ and $m^*$  denote the range of values for $m$. The two-step model selection method used herein can be summarized as follows:
\begin{enumerate}
  \item {Selection of $G$ and $m$}:
  Fit $G$ = $\{1, \hdots , G^*\}$ and $m $ = $ \{1, \hdots, m^*\}$ for the general ultrametric covariance structure FFFF. A total of $G^* \times m^* $ models are estimated. The $G$ and $m$ in the model combination ($G$, $m$, FFFF) that correspond to the maximized BIC are selected and used in Step~2.
  \item {Selection of covariance structure}: Fit all thirteen covariance structures for the selected $G$ and $m$ from Step~1. A total of 13 models are estimated. The covariance case in the model combination $(G, m, \text{case})$ that correspond to the maximized BIC is selected. 
\end{enumerate}
The selected model $(G, m, \text{case})$ consists of the $G$ and $m$ selected in Step~1 and the covariance case selected in Step~2. The two-step model selection is investigated in Simulation~1 (Section~\ref{sec:sim1}) and implemented on real datasets in Section~\ref{Sec5}.

\subsection{Stopping Rule}
Following \cite{bohning1994} and \cite{lindsay95}, a stopping rule based on Aitken's acceleration \citep{aitken1926} is used. At iteration $t$, Aitken's acceleration is defined as 
\begin{equation}
a^{(t)} = \frac{l^{(t+1)} - l^{(t)}}{l^{(t)} - l^{(t-1)}},
\end{equation}
where $l^{(t)}$ is the log-likelihood at iteration~$t$.
\cite{bohning1994} show that $a^{(t)}$ can be used to find an asymptotic estimate of the log-likelihood at iteration $t+1$ via
\begin{equation}
l^{(t+1)}_{\infty} = l^{(t)} + \frac{l^{(t+1)} - l^{(t)}}{1 - a^{(t)}}.
\end{equation}
Following \cite{mcnicholas2010}, the algorithm is stopped when
 $l^{(t+1)}_{\infty} - l^{(t)} \in (0, \epsilon).$ In the simulations and real data analyses herein, the tolerance value is set to $\epsilon = 10^{-4}$.

\section{Simulation Studies}
\label{Sec4}

\subsection{Overview} \label{sec:simover}
The proposed PUMMMs and the two-step model selection method are investigated in three simulation studies. The first simulation study investigates the effectiveness of the proposed two-step model selection method in comparison to the traditional use of the BIC. In the second simulation study, the performance of the PUMMM family is compared to the PUGMM family from the \texttt{PUGMM} package \citep{PUGMM} and Manly mixture models from the \texttt{ManlyMix} package \citep{ManlyMix} on simulated data generated from parsimonious ultrametric covariance structures. In the third simulation study, we compare the performance of the same methods used in Simulation 2, but on simulated data generated from covariance structures that are not inherently hierarchical. All methods are fit for $G = 1,\hdots, 5$, and the PUGMMs and PUMMMs are fit for $m = 1,\hdots, 5$. All methods are initialized using their respective {\sf R} function default settings. For the PUGMMs and the Manly mixture models, this is a $k$-means initialization. For the PUMMMs, this is a short EM using \texttt{ManlyMix}. The adjusted Rand index \citep[ARI;] []{hubert85} is used to assess clustering performance, where the ARI has an expected value of 0 under random classification and a value of 1 under perfect classification. 

\subsection{Simulation 1}\label{sec:sim1}

\begin{figure}[htp]
  \centering
  \includegraphics[width=0.98\textwidth]{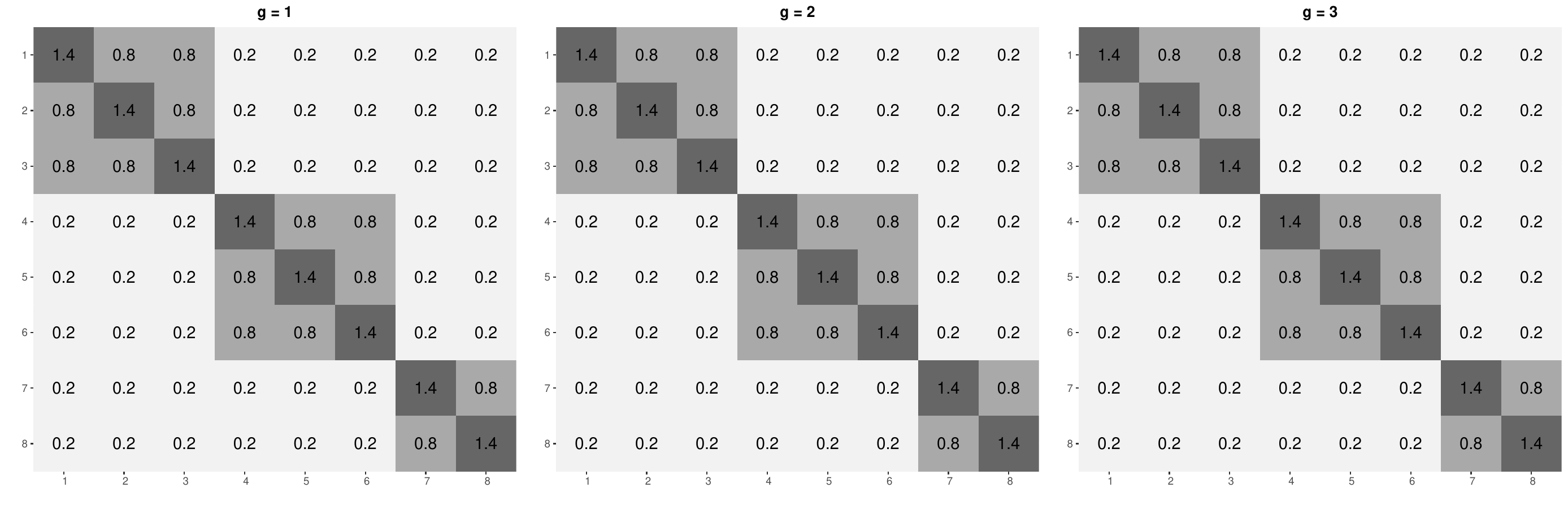}
   \vspace{-3.5mm}
  \caption{The EUUU covariance case used in Simulation 1.}
 \label{Fig1:sim1_EUUU}
  \includegraphics[width=0.98\textwidth]{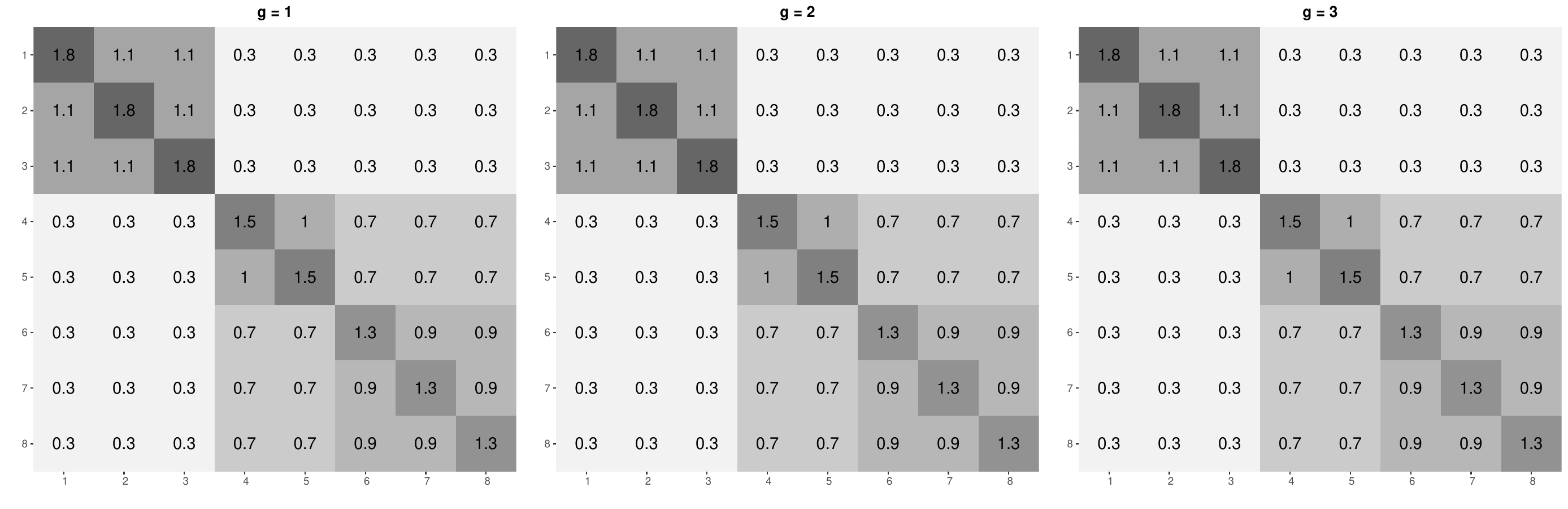}
  \vspace{-3.5mm}
  \caption{The EEEE covariance case used in Simulation 1.}
  \label{Fig2:sim1_EEEE}
  \includegraphics[width=0.98\textwidth]{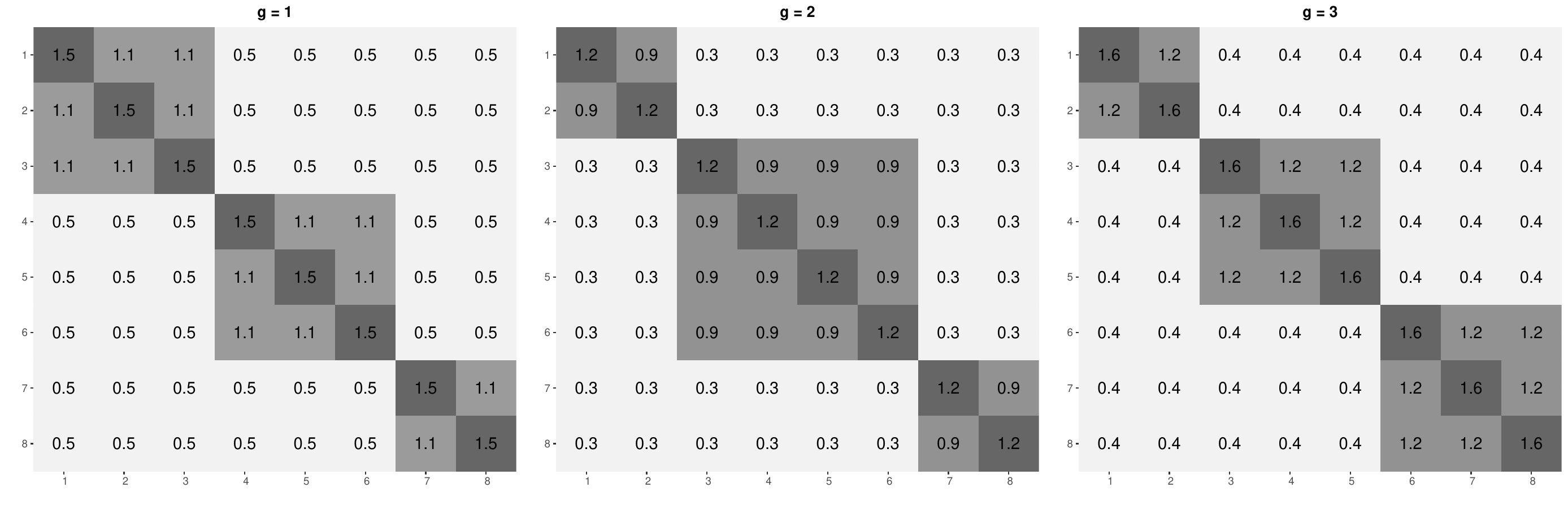}
  \vspace{-3.5mm}
  \caption{The FIII covariance case used in Simulation 1.}
  \label{Fig3:sim1_FIII}
  \includegraphics[width=0.98\textwidth]{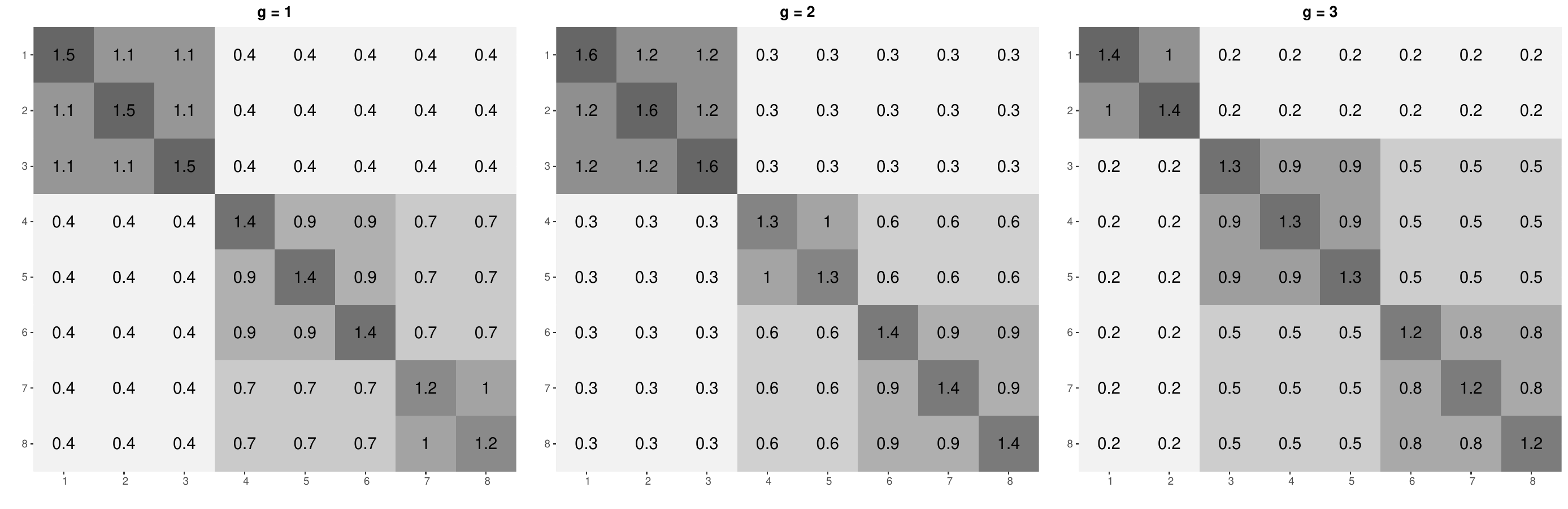}
  \vspace{-3.5mm}
  \caption{The FFFF covariance case used in Simulation 1.}
  \label{Fig4:sim1_FFFF}
\end{figure}

In this simulation study, a total of 160 samples are generated from a Manly mixture with $G = 3$, $p = 8$, $m = 3$, and $\pi_1 = \pi_2 = \pi_3 = 1/3$ for $n = 250$ and $n = 500$. Three transformation parameter settings and four covariance structures are considered for a total of 12 scenarios. The covariance structures include the most constrained covariance case EUUU, the least constrained case FFFF, as well as two moderately constrained cases, EEEE and FIII. These covariances are depicted in Figures~\ref{Fig1:sim1_EUUU}--\ref{Fig4:sim1_FFFF}.
The set of parameters for each scenario include one of the four covariances, one of the following transformations parameters,
 \begin{align*}
   \boldsymbol{\lambda}^{(1)} &= \begin{pmatrix}
        1.10 & 1.10 & 1.10 & 1.10 & 1.12 & 1.12 & 1.12 & 1.12 \\
       1.05 & 1.05 & 1.05 & 1.05 & 1.09 & 1.09 & 1.09 & 1.09 \\
       1.08 & 1.08 & 1.08 & 1.08 & 1.06 & 1.06 & 1.06 & 1.06 
       \end{pmatrix}, \\
       \boldsymbol{\lambda}^{(2)} &= \begin{pmatrix}
        2.10 & 2.10 & 2.10 & 2.10 & 2.12 & 2.12 & 2.12 & 2.12 \\
       2.05 & 2.05 & 2.05 & 2.05 & 2.09 & 2.09 & 2.09 & 2.09 \\
       2.08 & 2.08 & 2.08 & 2.08 & 2.06 & 2.06 & 2.06 & 2.06 
       \end{pmatrix}, \\
       \boldsymbol{\lambda}^{(3)} &= \begin{pmatrix}
        4.10 & 4.10 & 4.10 & 4.10 & 4.12 & 4.12 & 4.12 & 4.12 \\
       4.05 & 4.05 & 4.05 & 4.05 & 4.09 & 4.09 & 4.09 & 4.09 \\
       4.08 & 4.08 & 4.08 & 4.08 & 4.06 & 4.06 & 4.06 & 4.06 
       \end{pmatrix},
\end{align*}
 and 
\begin{equation*}
    \boldsymbol{\mu} =  \begin{pmatrix}
   10 & 10 & 10 & 10 & 15 & 15 & 15 & 15 \\ 
   15 & 15 & 15 & 15 & 25 & 25 & 25 & 25 \\
   20 & 20 & 20 & 20 & 20 & 20 & 20 & 20 \end{pmatrix},
\end{equation*} 
where each row in $\boldsymbol{\lambda}^{(k)}$, for $k = 1,2,3$, and $\boldsymbol{\mu}$ correspond to the $g$th cluster. A Gaussian extension of this simulation is given in Appendix \ref{app:sim1_ext}.

Table \ref{table:sim1} includes the percentage of times that the correct number of clusters $G$, variable groups $m$, covariance case, and model ($G$, $m$, case) are selected. The proposed two-step model selection consistently outperforms the BIC in selecting the correct $m$, case, and model, and both methods perform similarly in terms of selecting $G$. Looking more closely at the selection rate for $m$, the two-step method does about twice as well as the BIC for many of the scenarios. When investigating this result, the BIC tends to overestimate $m$, selecting $m$ = 4 or 5. A similar pattern emerges in the Gaussian setting (Appendix~\ref{app:sim1_ext}). For the EEEE and FFFF scenarios, the detection rate for the models are lower as both the BIC and two-step model selection approaches tend to select the more parsimonious models EUUE and FIIF, respectively.

\begin{sidewaystable}[p]
    \centering
	\caption{A comparison of model selection approaches for the PUMMMs: the proposed two-step method (left) and the BIC (right). The table summarizes the amount of times that the correct $G$, $m$, case, and model ($G$, $m$, case) are selected.}	
	\scalebox{1}{
		\begin{tabular*}{0.98\textwidth}{@{\extracolsep{\fill}}ccccccccccccc}
		  \noalign{\smallskip}\hline
 		&&&\multicolumn{2}{c}{$G$} &\multicolumn{2}{c}{$m$} &\multicolumn{2}{c}{Case} & \multicolumn{2}{c}{($G$, $m$, case)} \\ 
\cline{4-5}\cline{6-7}\cline{8-9}\cline{10-11}\cline{12-13}
		              $\boldsymbol{\lambda}^{(k)}$ & Model& $n$ &Two-step&BIC&Two-step&BIC&Two-step&BIC&Two-step&BIC\\
\hline
\multirow{8}{*}{$\boldsymbol{\lambda}^{(1)}$} & \multirow{2}{*}{EUUU}&$250$& 98.67\% & 100.00\% & 82.67\% & 43.33\% & 77.33\% & 42.00\% & 76.67\% & 42.00\% \\
& &$500$ & 96.00\% & 100.00\% & 82.67\% & 46.67\% & 80.67\% & 46.67\% & 78.67\% & 46.67\% \\
& \multirow{2}{*}{EEEE}&$250$& 100.00\% & 100.00\% & 85.33\% & 13.33\% & 18.00\% & 4.00\% & 18.00\% & 4.00\% \\
& &$500$ & 100.00\% & 100.00\% & 87.33\% & 27.33\%  & 48.67\% & 18.67\% & 48.67\% & 18.67\% \\
& \multirow{2}{*}{FIII}&$250$& 96.67\% & 100.00\% & 98.00\% & 66.67\% & 86.00\% & 63.33\% & 82.67\% & 63.33\% \\
& &$500$ & 100.00\% & 100.00\% & 100.00\% & 70.67\%  & 89.33\% & 69.33\% & 89.33\% & 69.33\% \\
& \multirow{2}{*}{FFFF}&$250$& 100.00\% & 100.00\% & 95.33\% & 31.33\% & 3.33\% & 0.00\% & 3.33\% & 0.00\% \\
& &$500$ & 98.67\% & 100.00\% & 94.00\% & 50.00\%  & 38.67\% & 20.67\% & 38.67\% & 20.67\% \\
\midrule
\multirow{8}{*}{$\boldsymbol{\lambda}^{(2)}$}& \multirow{2}{*}{EUUU}&$250$& 99.33\% & 100.00\% & 86.00\% & 45.33\%  & 82.00\% & 44.67\% & 82.00\% & 44.67\% \\
& &$500$& 98.00\% & 100.00\% & 88.67\% & 50.67\% & 88.00\% & 50.67\% & 86.67\% & 50.67\% \\
& \multirow{2}{*}{EEEE}&$250$& 100.00\% & 99.33\% & 84.00\% & 16.00\%  & 22.00\% & 6.00\% & 22.00\% & 6.00\% \\
& &$500$ & 100.00\% & 100.00\% & 86.67\% & 34.67\%  & 46.00\% & 24.67\% & 46.00\% & 24.67\% \\
& \multirow{2}{*}{FIII}&$250$& 97.33\% & 100.00\% & 96.67\% & 64.00\% & 84.00\% & 62.00\% & 81.33\%  & 62.00\% \\
& &$500$ & 100.00\% & 100.00\% & 100.00\% & 73.33\%  & 87.33\% & 70.67\% & 87.33\% & 70.67\% \\
& \multirow{2}{*}{FFFF}&$250$& 98.67\% & 100.00\% & 96.00\% & 28.67\% & 4.67\% & 0.67\% & 4.67\% & 0.67\% \\
& &$500$ & 100.00\% & 100.00\% & 98.00\% & 45.33\%  & 50.00\% & 27.33\% & 49.33\% & 27.33\% \\
\midrule
\multirow{8}{*}{$\boldsymbol{\lambda}^{(3)}$} & \multirow{2}{*}{EUUU}&$250$& 98.67\% & 100.00\% & 91.33\% & 44.67\% & 88.67\% & 43.33\% & 88.00\% & 43.00\% \\
& &$500$ & 98.67\% & 100.00\% & 89.33\% & 48.67\% & 89.33\% & 48.67\% & 88.00\% & 48.67\% \\
& \multirow{2}{*}{EEEE}&$250$& 100.00\% & 100.00\% & 84.67\% & 18.00\%  & 22.67\% & 8.67\% & 22.67\% & 8.67\% \\
& & $500$ & 100.00\% & 100.00\% & 91.33\% & 30.67\%  & 52.00\% & 22.00\% & 52.00\% & 22.00\% \\
& \multirow{2}{*}{FIII}&$250$& 96.67\% & 100.00\% & 96.67\% & 69.33\% & 86.67\% & 67.33\% & 83.33\% & 67.33\% \\
& &$500$ & 100.00\% & 100.00\% & 100.00\% & 77.33\%  & 87.33\% & 72.00\% & 87.33\% & 72.00\% \\
& \multirow{2}{*}{FFFF}&$250$& 99.33\% & 100.00\% & 98.00\% & 26.67\% & 8.00\% & 1.33\% & 8.00\% & 1.33\% \\
& &$500$ & 96.00\% & 100.00\% & 98.00\% & 42.67\%  & 55.33\% & 26.67\% & 54.00\% & 26.67\% \\
\hline
	\end{tabular*}}
	\label{table:sim1}
\end{sidewaystable}

\subsection{Simulation 2}
\begin{figure}[t]
  \centering
  \includegraphics[width=1.0\textwidth]{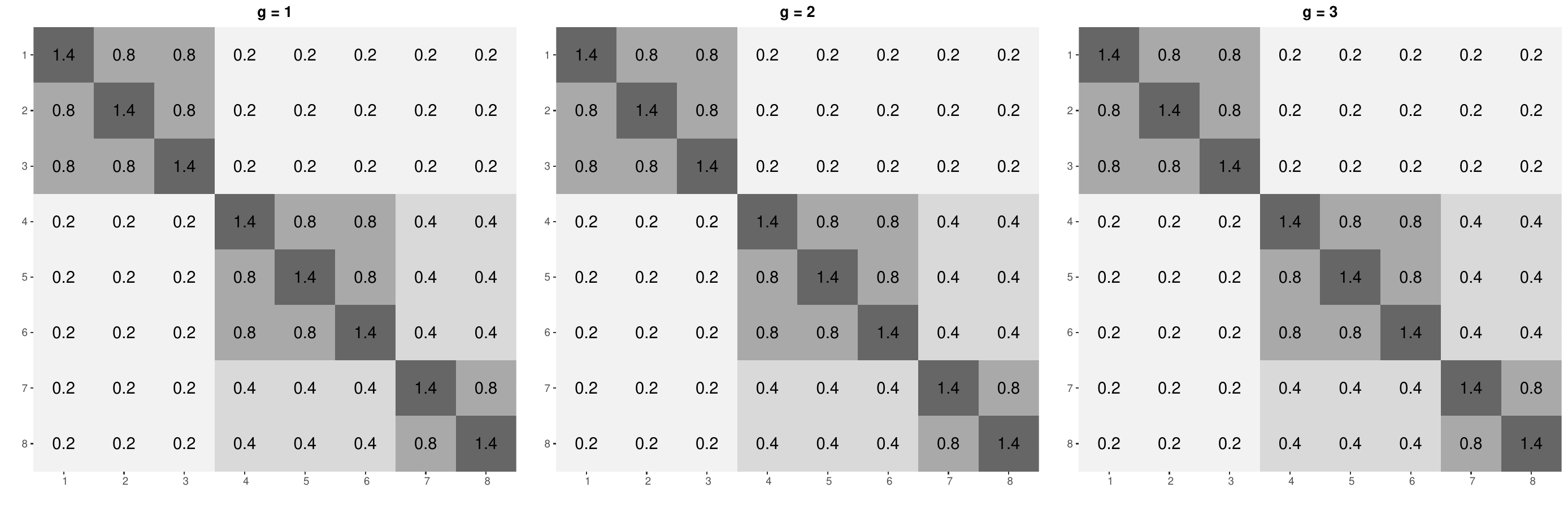}
  \vspace{-3.5mm}
  \caption{The EUUE covariance case used in Simulation 2.}
  \label{Fig:sim2_EUUE}
  \includegraphics[width=1.0\textwidth]{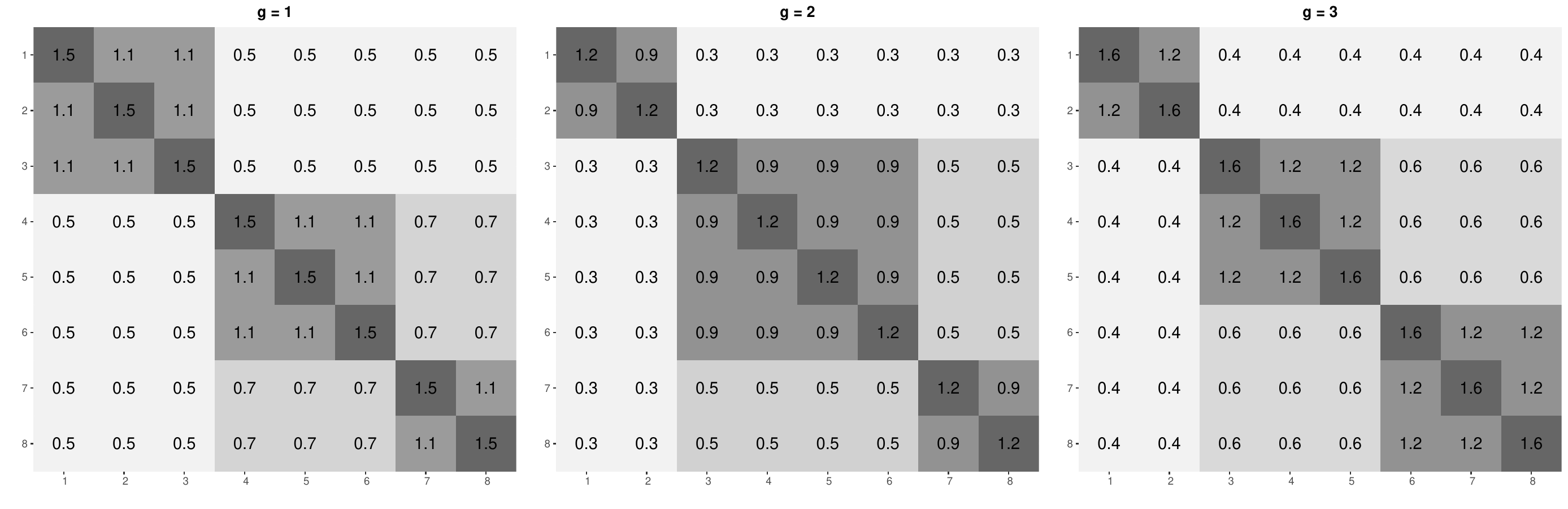}
  \vspace{-3.5mm}
  \caption{The FIIF covariance case used in Simulation 2.}
  \label{Fig:sim2_FIIF}
\end{figure}

This simulation compares the clustering performance of the PUMMM family to the PUGMM family, a full Manly mixture, a Manly mixture with forward model selection, and a Manly mixture with backward model selection. For the PUMMMs and the PUGMMs, we report the clustering performance for both the two-step model selection method and the BIC. To compare the methods on skewed data generated from ultrametric structures, we considered Manly and skew-$t$ scenarios for the EUUE and FIIF cases. The covariances are illustrated in Figures \ref{Fig:sim2_EUUE} and \ref{Fig:sim2_FIIF}, respectively. Both scenarios are generated from a mixture with $G = 3$, $p = 8$, $m = 3$, $n = 300$, and $\pi_1 = \pi_2 = \pi_3 = 1/3$ for a total of 160 samples. The parameters for each scenario include either the EUUE and FIIF covariance structure, and
\begin{equation*}
\boldsymbol{\mu}  =  \begin{pmatrix}
   13 & 13 & 12 & 12 & 16 & 16 & 17 & 17 \\
   15 & 14 & 14 & 15 & 21 & 20 & 20 & 21 \\
   18 & 17 & 18 & 17 & 19 & 18 & 19 & 18 \end{pmatrix}.
\end{equation*} For the Manly scenario, the transformation parameter is \begin{equation*}
\boldsymbol{\lambda} = \begin{pmatrix}
  2.10 & 2.10 & 2.10 & 2.10 & 2.10 & 2.10 & 2.10 & 2.10 \\
  2.12 & 2.12 & 2.12 & 2.12 & 2.12 & 2.12 & 2.12 & 2.12 \\
  2.08 & 2.08 & 2.08 & 2.08 & 2.08 & 2.08 & 2.08 & 2.08
\end{pmatrix}.
\end{equation*} 
For the skew-$t$ scenario, the skewness parameter is 
 \begin{equation*}
\boldsymbol{\alpha} = \begin{pmatrix}
  6 & 6 & 6 & 6 & 6 & 6 & 6 & 6 \\
  8 & 8 & 8 & 8 & 8 & 8 & 8 & 8 \\
  4 & 4 & 4 & 4 & 4 & 4 & 4 & 4
\end{pmatrix}, 
\end{equation*}
with degrees of freedom $\nu_1 = 10$, $\nu_2 = 20$, and $\nu_3 = 15$,  where each row in $\boldsymbol{\alpha}$ corresponds to the $g$th cluster.

From Table \ref{table:sim2}, we can see the PUMMMs out perform the Manly mixture models in the EUUE settings, but tend to perform equivalently in the FIIF settings. Additionally, the ARI is the same for the PUMMMs with the two-step model selection as for the PUMMMs with the BIC model selection; the same occurs for the PUGMMs. The BIC is larger for the PUMMMs and PUGMMs with the BIC model selection than with the two-step model selection. Expected, as a correct model selection by the BIC would eliminate the need for the proposed two-step model selection method.
\begin{table*}[!ht]
	\centering
 \small
	\caption{A comparison of BIC and ARI values, respectively, for the PUMMMs, the PUGMMs, and Manly mixture models on simulated data as described in Simulation 2.}	
	\scalebox{1}{
		\begin{tabular*}{1.0\textwidth}{@{\extracolsep{\fill}}lllccccccccc}
		  \noalign{\smallskip}\toprule
 		&&&\multicolumn{2}{c}{BIC} &\multicolumn{2}{c}{ARI}\\ 
\cline{4-5}\cline{6-7}
Scenario & Model & Method & Mean & Std.~Dev. & Mean & Std.~Dev.\\
\midrule
\multirow{7}{*}{Manly} & \multirow{7}{*}{EUUE} & PUMMMs (Two-step) & -2924.86 &  1134.89 & 0.98 & 0.02\\
& & PUMMMs (BIC) & -2921.52 & 1134.13 & 0.98 & 0.02\\
& & PUGMMs (Two-step) & -3082.63 & 989.81 & 0.94 & 0.10\\
& & PUGMMs (BIC)  &  -3070.59 & 990.86 & 0.94 & 0.09\\
& & Manly Mixture (Full) & -3423.89 & 995.532 & 0.71 & 0.34\\
& & Manly Mixture (Forward) &  -3324.49& 1034.83 & 0.89 & 0.18\\
& & Manly Mixture (Backward) & -3332.47 & 1043.12 & 0.71& 0.34\\
\midrule
\multirow{7}{*}{Manly} & \multirow{7}{*}{FIIF}& PUMMMs (Two-step) & -3432.52 &  113.54 & 0.97 & 0.04\\
& & PUMMMs (BIC)  & -3420.58 & 108.87 & 0.97 & 0.02\\
& & PUGMMs (Two-step) & -3423.79 & 107.00 & 0.97 & 0.03\\
& & PUGMMs (BIC) & -3421.05 & 105.71 & 0.97 & 0.02\\
& & Manly Mixture (Full)  & -3721.69 & 102.14 & 0.94 &  0.12\\
& & Manly Mixture (Forward) & -3630.50 & 102.67 & 0.97 & 0.05\\
& & Manly Mixture (Backward) & -3628.61 & 101.68 & 0.94 & 0.12\\
\midrule
\multirow{7}{*}{Skew-$t$} & \multirow{7}{*}{EUUE}& PUMMMs (Two-step) & -4231.757 & 88.59 & 0.98 & 0.013\\
& & PUMMMs (BIC)  & -4187.723 & 387.364 & 0.97 & 0.049\\
& & PUGMMs (Two-step) & -4164.764 & 87.315 & 0.98 & 0.013\\
& & PUGMMs (BIC) & -4153.125 &  88.285 & 0.98 & 0.024\\
& & Manly Mixture (Full) & -4571.288 & 83.556 & 0.56 & 0.067 \\
& & Manly Mixture (Forward) & -4499.962 & 84.177 & 0.64 & 0.173\\
& & Manly Mixture (Backward) &  -4501.735 & 84.139 & 0.56 & 0.067\\
\midrule
\multirow{7}{*}{Skew-$t$} & \multirow{7}{*}{FIIF}& PUMMMs (Two-step) & -3721.742 & 109.194 & 0.99 & 0.012\\
& & PUMMMs (BIC)  & -3716.351 & 109.964 & 0.99 & 0.012\\
& & PUGMMs (Two-step)  & -3691.976 & 113.217 & 0.98 & 0.017\\
& & PUGMMs (BIC) & -3687.206 & 113.902 & 0.98 & 0.025\\
& & Manly Mixture (Full)  & -4019.175 & 111.328 & 0.98 &  0.016\\
& & Manly Mixture (Forward) & -3915.323 & 110.217 & 0.98 & 0.016\\
& & Manly Mixture (Backward) & -3915.178 & 110.336 & 0.98 & 0.016\\
\bottomrule
	\end{tabular*}}
	\label{table:sim2}
\end{table*}

\subsection{Simulation 3}
\begin{figure}[t]
  \centering
  \includegraphics[width=0.90\textwidth]{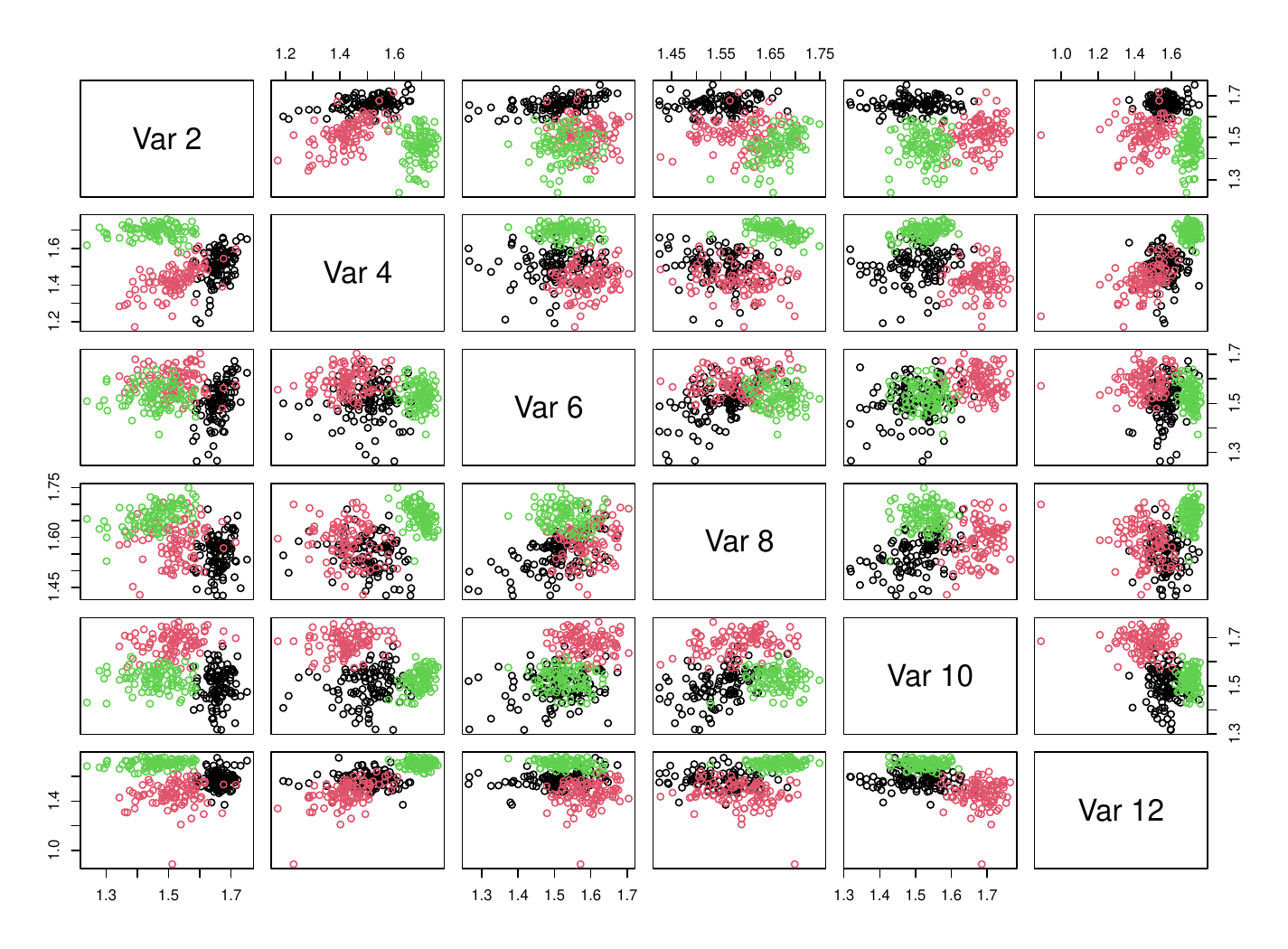}
  \caption{An example of one of the Manly simulated datasets in Simulation 3.}
  \label{Fig:sim3_manly}
\end{figure}

In this third simulation, the performance of the PUMMMs and the methods described in Simulation 2 are compared on simulated data where the covariance structure is non-hierarchical. Gaussian, Manly, and skew-$t$ scenarios are considered where each scenario is generated from their respective mixture model with 
$G = 3$, $p = 12$, $n = 300$, and $\pi_1 = \pi_2 = \pi_3 = 1/3$ for a total of 160 samples. The covariance matrices are generated from a positive-definite covariance matrix generator in {\sf R}. The component means are sampled with replacement from the sets $\{15, \hdots 25\}$, $\{10, \hdots 25\}$, $\{15, \hdots 30\}$, for each respective cluster. A seed is set to guarantee that the same covariances and means in each sample. For the Manly mixture, the transformation parameter is
\begin{equation*}
\boldsymbol{\lambda} = \begin{pmatrix}
  2.50 & 2.50 & 2.50 & 2.50 & 2.50 & 2.50 & 2.50 & 2.50 & 2.50 & 2.50 & 2.50 \\
  2.45 & 2.45 & 2.45 & 2.45 & 2.45 & 2.45 & 2.45 & 2.45 & 2.45 & 2.45 & 2.45 \\
  2.55 & 2.55 & 2.55 & 2.55 & 2.55 & 2.55 & 2.55 & 2.55 & 2.55 & 2.55 & 2.55 
\end{pmatrix}.
\end{equation*}
A visualization of the Manly scenario is given in Figure \ref{Fig:sim3_manly}. For the skew-$t$ scenario, the skewness parameter $\boldsymbol{\alpha}$ is 
\begin{equation*}
\boldsymbol{\alpha} = \begin{pmatrix}
  2 & 2 & 2 & 2 & 2 & 2 & 2 & 2 & 2 & 2 & 2 \\
  5 & 5 & 5 & 5 & 5 & 5 & 5 & 5 & 5 & 5 & 5 \\
  3 & 3 & 3 & 3 & 3 & 3 & 3 & 3 & 3 & 3 & 3
\end{pmatrix}, 
\end{equation*}
with degrees of freedom $\nu_1=10$, $\nu_2=20$, and $\nu_3=30$.

In Table \ref{table:sim3}, we can see that all methods perform well, but the three Manly mixture model variants perform the best with a mean ARI of 1.00. In the Gaussian and Manly scenarios, the PUMMMs and PUGMMs with the two-step model selection perform marginally better than with the BIC; however, in the skew-$t$ scenario, the two-step model selection outperforms the BIC with a difference of about 0.1 in ARI. Overall, the PUMMMs perform comparably to the Manly mixture models in \texttt{ManlyMix}.

\begin{table*}[!ht]
	\centering
 \small
	\caption{A comparison of the BIC and ARI for the PUMMMs, the PUGMMs, and Manly mixture models on simulated data as described in Simulation 3.}	
	\scalebox{1}{
		\begin{tabular*}{1.0\textwidth}{@{\extracolsep{\fill}}llcccccccccc}
		  \noalign{\smallskip}\toprule
 		&&\multicolumn{2}{c}{BIC} &\multicolumn{2}{c}{ARI}\\ 
\cline{3-4}\cline{5-6}
Scenario & Method & mean & sd & mean & sd\\
\midrule
\multirow{7}{*}{Gaussian}& PUMMMs (Two-step) & -6544.29 & 89.923 & 0.99 & 0.026\\
& PUMMMs (BIC) & -6452.57 & 199.021 & 0.98 & 0.055\\
& PUGMMs (Two-step) & -6476.99 & 90.283 & 0.93 & 0.094\\
& PUGMMs (BIC)& -6405.35 & 87.907 & 0.90 & 0.109\\
& Manly Mixture Model (Full) & -5100.61 & 85.381 & 1.00 & 0.000\\
& Manly Mixture Model (Backward) & -4933.14 & 85.046 & 1.00 & 0.000\\
& Manly Mixture Model (Forward)  & -4933.14 & 85.045 & 1.00 & 0.000\\
\midrule
\multirow{7}{*}{Manly }& PUMMMs (Two-step) & -6644.60 & 105.359 & 1.00 & 0.020\\
& PUMMMs (BIC) & -6612.14 & 152.502 & 0.98 & 0.051\\
& PUGMMs (Two-step) & -6724.75 & 109.202 & 0.95 & 0.084\\
& PUGMMs (BIC) & -6680.35 & 148.258 & 0.95 & 0.085\\
& Manly Mixture Model (Full) & -5318.54 & 118.889 & 1.00 & 0.000\\
& Manly Mixture Model (Forward) & -5264.11 & 117.219 & 1.00 & 0.000\\
& Manly Mixture Model (Backward) & -5264.08 & 117.243 & 1.00 & 0.000\\
\midrule
\multirow{7}{*}{Skew-$t$}& PUMMMs (Two-step)  & -6560.78 & 659.987 & 0.99 & 0.036 \\
& PUMMMs (BIC)  & -6113.64 & 1639.883 & 0.90 & 0.105\\
& PUGMMs (Two-step) & -6466.06 & 764.356 & 0.94 & 0.085\\
& PUGMMs (BIC) & -5406.27 & 2192.417 & 0.83 & 0.147\\
& Manly Mixture Model (Full) & -5311.5 & 99.742 & 1.00 & 0.004\\
& Manly Mixture Model (Forward) & -5148.14 & 96.442 & 1.00 & 0.002\\
& Manly Mixture Model (Backward) & -5148.14 & 96.502 & 1.00 & 0.002\\
\bottomrule
	\end{tabular*}}
	\label{table:sim3}
\end{table*}

\section{Real Data Results}
\label{Sec5}
In this section, the PUMMMs are evaluated on real data including several benchmark datasets in Section~\ref{Benchmark}, where the labels are known a priori, as well as on a Harbour Metals dataset \citep{roberts08} in \ref{Harbourmetals} which contains the concentration levels of several metals in two seaweed species. 
The initializations for each method are as described in Section~\ref{sec:simover} and clustering performance is assessed using the ARI.

\subsection{Benchmark datasets}
\label{Benchmark}

We compare the PUMMMs to the PUGMMs, a full Manly mixture, a Manly mixture with forward model selection, and a Manly mixture with backward model selection on benchmark datasets seen in both the Gaussian and non-Gaussian clustering literature. Table~\ref{table:realdata_info} provides feature information for the datasets, and the source information is in Table~\ref{table:realdata_source}. The two-step model selection method and the BIC are implemented for the PUMMMs and PUGMMs, resulting in a total of seven methods. All methods are fit for $G = 1, \hdots, G^*+2$, where $G^*$ is the theoretical number of clusters. If $G^*$ is not selected, the method is refit with $G = G^*$, and both results are recorded with the latter in parentheses. The PUMMMs and PUGMMs are fit over
$$m=\begin{cases}
1, \hdots, p & \text{for } p < 10,\\
1, \hdots ,10 & \text{otherwise.}
\end{cases}$$
Details of the selected models, the number of free covariance parameters $\rho_{\text{cov}}$, and the ARI are reported in Table~\ref{tab:realdata}. 
\begin{table}[!htb]
\centering
\caption{The features and source information for the datasets used in Table \ref{tab:realdata}.}
\begin{tabular}{l c c c l l} 
 \toprule
 dataset & $n$ & $p$ & $G^*$ & Source \\[0.5ex] 
 \hline
AIS & 202 & 11 & 2 & \texttt{ManlyMix} \\
Banknote & 200 & 6 & 2 & \texttt{mclust} \\
Body & 507 & 24 & 2 & \texttt{gclus} \\
Diabetes & 145 & 3 & 3 & \texttt{mclust} \\
Olive & 572 & 8 & 3 & \texttt{pgmm} \\
Thyroid & 215 & 5 & 3 & \texttt{mclust} \\
Wine 13 & 178 & 13 & 3 & \texttt{ContaminatedMixt} \\
Wine 27 & 178 & 27 & 3 & \texttt{pgmm} \\
 \bottomrule
\end{tabular}
\label{table:realdata_info}
\end{table}

Table \ref{tab:realdata} shows the strong performance of the PUMMMs. For the four datasets where the two-step model selection correctly identifies $G$ --- i.e., AIS, Thyroid, Wine~13, and Wine~27 --- the ARIs are high. For the two high-dimensional datasets, Body and Wine~27, the PUMMMs significantly decrease the number of free covariance parameters. In particular, for the Body dataset with $G = 2$, the number of free covariance parameters is reduced from 600 to 37 while achieving the highest ARI of 0.95. Compared to the PUGMMs, the PUMMMs achieve an improved ARI for several datasets, notably Wine~13 and Wine~27. Compared to the Manly mixture models, the PUMMMs achieve comparable clustering results for Banknote, Olive, Wine~13, and Wine~27, and an improved clustering performance for AIS, Body, and Diabetes. 

\begin{sidewaystable}[p]
\centering
\small
\caption{Comparison of results across selected methods. Values in parentheses denote when $G^*$ has been specified.}
\begin{tabular*}{0.95\textwidth}{@{\extracolsep{\fill}}l ccccc ccccc}
\toprule
& \multicolumn{5}{c}{PUMMM (Two-step)} & \multicolumn{5}{c}{PUMMM} \\
\cmidrule(lr){2-6} \cmidrule(lr){7-11}
Dataset & $G$ & $m$ & Case & $\rho_{\text{cov}}$ & ARI 
        & $G$ & $m$ & Case & $\rho_{\text{cov}}$ & ARI \\
\midrule
Ais       & 2 & 7 & EEEE & 30 & 0.94 
          & 4(2) & 10(9) & FIIF(EUUE) & 84(36) & 0.58(0.92) \\

Banknote  & 3(2) & 5(5) & EEEE(EEEE) & 20(20) & 0.85(0.98) 
          & 4(2) & 6(5) & EUUE(EEEE) & 13(20) & 0.55(0.98) \\

Body & 4(2) & 9(6) & EEEE(EEEU) & 50(37) & 0.52(0.95) 
          & 4(2) & 9(9) & EEEE(EUUE) & 50(34) & 0.52(0.92) \\

Diabetes  & 2(3) & 2(2) & FFFI(FFFI) & 16(24) & 0.46(0.73) 
          & 2(3) & 2(2) & FFFF(FFFI) & 16(24) & 0.47(0.77) \\

Olive     & 4(3) & 4(3) & FFFF(FFFF) & 76(48) & 0.69(1.00) 
          & 2(3) & 4(3) & EEEE(FFFF) & 19(48) & 0.82(1.00) \\

Thyroid   & 3 & 2 & FFFI & 30 & 0.89 
          & 3 & 4 & EEEE & 16 & 0.83 \\

Wine 13   & 3 & 8 & FIIF & 66 & 0.98 
          & 3 & 9 & FIIF & 69 & 0.95 \\

Wine 27   & 3 & 8 & FFFF & 150 & 0.91 
          & 3 & 10 & FFFI & 144 & 0.90 \\
\midrule
\end{tabular*}

\vspace{2mm}

\begin{tabular*}{0.95\textwidth}{@{\extracolsep{\fill}}l ccccc ccccc}
\toprule
& \multicolumn{5}{c}{PUGMM (Two-step)} & \multicolumn{5}{c}{PUGMM} \\
\cmidrule(lr){2-6} \cmidrule(lr){7-11}
Dataset & $G$ & $m$ & Case & $\rho_{\text{cov}}$ & ARI 
        & $G$ & $m$ & Case & $\rho_{\text{cov}}$ & ARI \\
\midrule
Ais       & 3(2) & 6(6) & EEEE(EEEE) & 27(27) & 0.62(0.88) 
          & 3(2) & 10(8) & FIIF(EEEE) & 63(33) & 0.51(0.90) \\

Banknote  & 2 & 4 & EEEE & 17 & 0.98 
          & 4(2) & 5(5) & EEEE(EEEE) & 20(20) & 0.56(0.96) \\

Body & 4(2) & 8(6) & EEEE(FFFI) & 47(74) & 0.50(0.61) 
          & 4(2) & 7(7) & EEEE(FFFI) & 44(78) & 0.46(0.60) \\
          
Diabetes  & 5(3) & 2(2) & FFFI(FFFI) & 40(24) & 0.68(0.51) 
          & 5(3) & 2(2) & FFFI(FFFI) & 40(24) & 0.68(0.51) \\

Olive     & 5(3) & 6(5) & FFFI(FFFF) & 95(66) & 0.38(0.47) 
          & 5(3) & 6(8) & FFFI(FIIF) & 105(66) & 0.38(0.53) \\

Thyroid   & 3 & 2 & FFFI & 30 & 0.88 
          & 3 & 3 & FFFI & 36 & 0.85 \\

Wine 13   & 3 & 8 & FFFF & 108 & 0.93 
          & 3 & 8 & FFFF & 108 & 0.93 \\

Wine 27   & 3 & 9 & FFFI & 138 & 0.88 
          & 3 & 9 & FFFI & 138 & 0.88 \\
\midrule
\end{tabular*}

\vspace{2mm}

\begin{tabular*}{0.95\textwidth}{@{\extracolsep{\fill}}l ccc ccc ccc}
\toprule
& \multicolumn{3}{c}{ManlyMix (Full)} 
& \multicolumn{3}{c}{ManlyMix (Forward)} 
& \multicolumn{3}{c}{ManlyMix (Backward)} \\
\cmidrule(lr){2-4} \cmidrule(lr){5-7} \cmidrule(lr){8-10}
Dataset & $G$ & $\rho_{\text{cov}}$ & ARI 
        & $G$ & $\rho_{\text{cov}}$ & ARI 
        & $G$ & $\rho_{\text{cov}}$ & ARI \\
\midrule
Ais       & 2 & 110 & 0.72 
          & -- & -- & -- 
          & 2 & 110 & 0.69 \\

Banknote  & 3(2) & 63(42) & 0.84(0.98) 
          & 3(2) & 63(42) & 0.84(0.98) 
          & 3(2) & 63(42) & 0.85(0.98) \\

Body      & 1(2) & 300(600) & 0.00(0.86) & 1(2) &
          300(600) & 0.00(0.88) & 1(2) & 300(600) & 0.00(0.87) \\

Diabetes  & 2(3) & 12(18) & 0.39(0.55) 
          & 3 & 18 & 0.70 
          & 2(3) & 12(18) & 0.46(0.61) \\

Olive     & 5(3) & 180(108) & 0.61(1.00) 
          & 5(3) & 180(108) & 0.63(1.00) 
          & 5(3) & 180(108) & 0.62(1.00) \\

Thyroid   & 3 & 45 & 0.70 
          & 3 & 45 & 0.86 
          & 3 & 45 & 0.86 \\

Wine 13   & --(3) & --(273) & --(0.95) 
          & 2(3) & 182(273) & 0.45(0.95) 
          & --(3) & --(273) & --(0.95) \\

Wine 27   & --(3) & --(1134) & --(0.95) 
          & --(3) & --(1134) & --(0.96) 
          & --(3) & --(1134) & --(0.96) \\
\bottomrule
\end{tabular*}
\label{tab:realdata}
\end{sidewaystable}

\begin{figure}[!ht]
\centering
\begin{subfigure}[b]{0.46\textwidth}
    \centering
    \includegraphics[width=\textwidth]{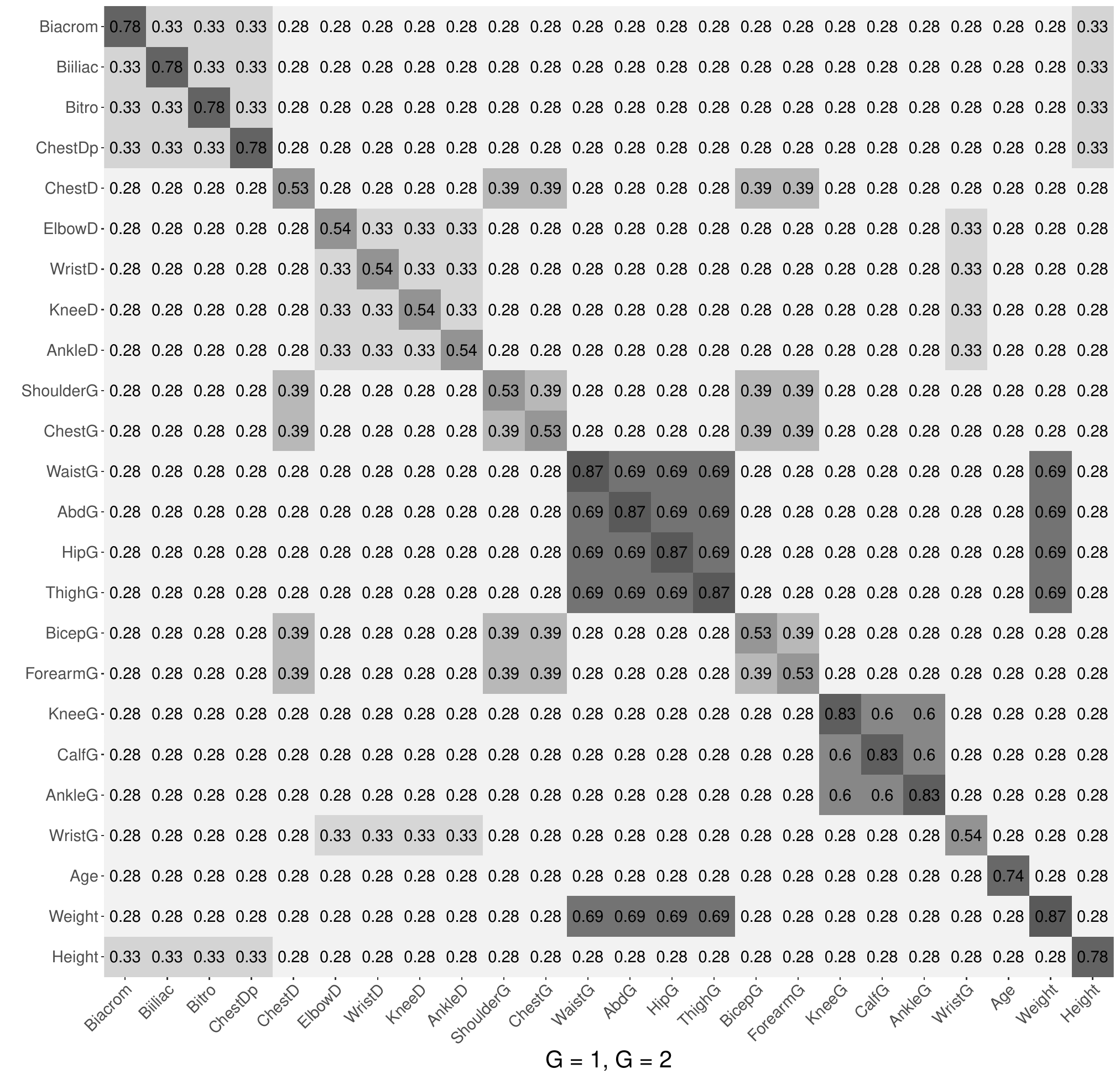}
    \caption{Heat Map}
    \label{fig:body_cov}
\end{subfigure}
\hfill
\begin{subfigure}[b]{0.53\textwidth}
    \centering
    \includegraphics[width=\textwidth]{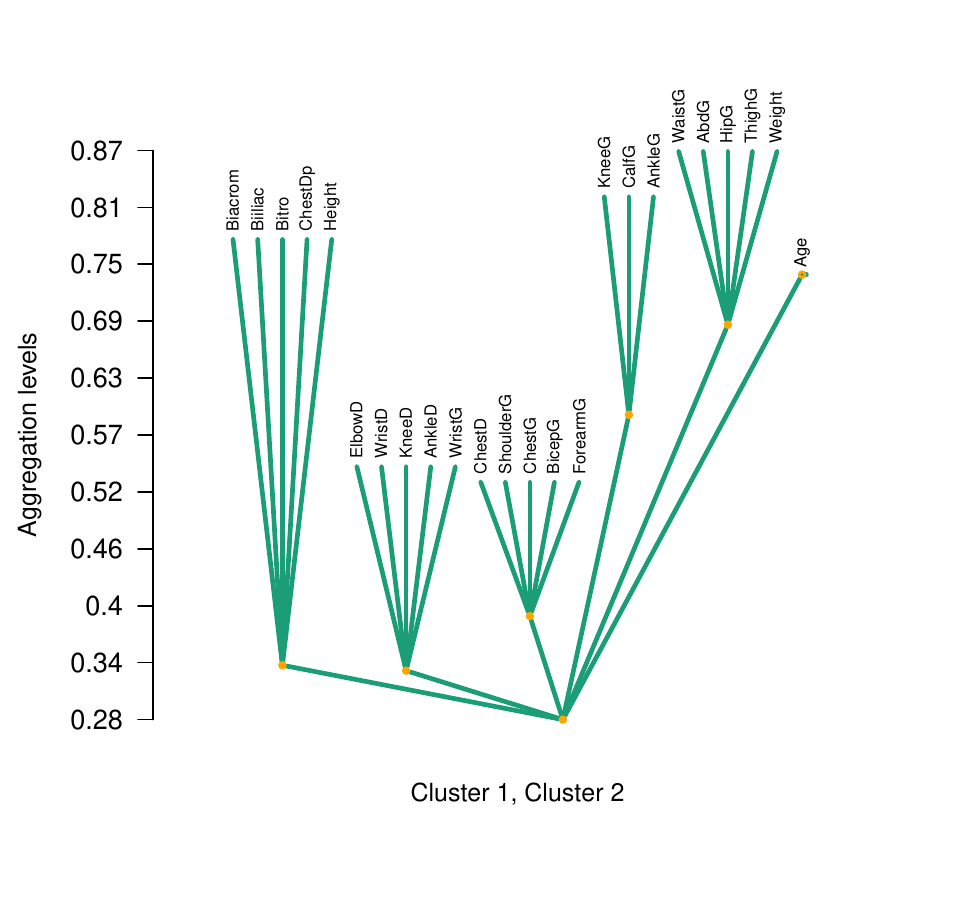}
    \caption{Path Diagram}
    \label{fig:body_path}
\end{subfigure}
\caption{Covariance heatmap (a) and path diagram (b) corresponding to the PUMMM results for the Body dataset.}
\end{figure}

The successful performance of the PUMMMs indicate that this class of models can introduce parsimony to Manly mixture models with the potential to identify hierarchical relationships within the clusters. As described in Section \ref{Sec21}, the main diagonal corresponds to the $\boldsymbol{\Sigma}_{V_g}$, the off-diagonal block matrices correspond to $\boldsymbol{\Sigma}_{W_g}$, and the remaining elements correspond to $\boldsymbol{\Sigma}_{B_g}$. The covariance values are directly associated with the aggregation levels, which are a measure of similarity, or concordance, among variables. To demonstrate this hierarchy, we include a heatmap of the covariance (Figure \ref{fig:body_cov}) and corresponding path diagram (Figure \ref{fig:body_path}) for the Body dataset when $G = 2$. The PUMMMs with the two-step model selection selected $m = 6$ and the EEEU covariance structure. A description for each variable in the Body dataset is given in the \texttt{gclus} package. 

Referencing Figure \ref{fig:body_path}, all groups, including the singleton Age, are aggregated at a low level of covariance, indicating that these six latent groups distinctly contribute to the covariance. The group consisting of WaistG, AbdG, HipG, ThightG and Weight exhibit a large within-group covariance of 0.69, a strong internal concordance. This is consistent with domain expectations, as overall body mass is closely related to these measurements, whereas variables such as knee or elbow measurements are less directly associated. The remaining groups exhibit anatomical regions. For example, one group is formed by KneeG, CalfG, and AnkleG, with another formed by ChestD, ShouldersG, ChestG, BicepsG, and ForearmG.

The observations in clusters 1 and 2 correspond to females and males, respectively. Both clusters share the same hierarchical structure, indicating that the same variable groupings and levels of concordance characterize each cluster. This suggests that, despite potential physiological differences, the underlying relationships among these measurements remain consistent across the clusters, even if their distributions differ in $\boldsymbol{\mu}_g$ and $\boldsymbol{\lambda}_g$.

\subsection{Harbour Metals}
\label{Harbourmetals}
\begin{figure}[!ht]
\centering
\begin{subfigure}[b]{0.5\textwidth}
    \centering
    \includegraphics[width=\textwidth]{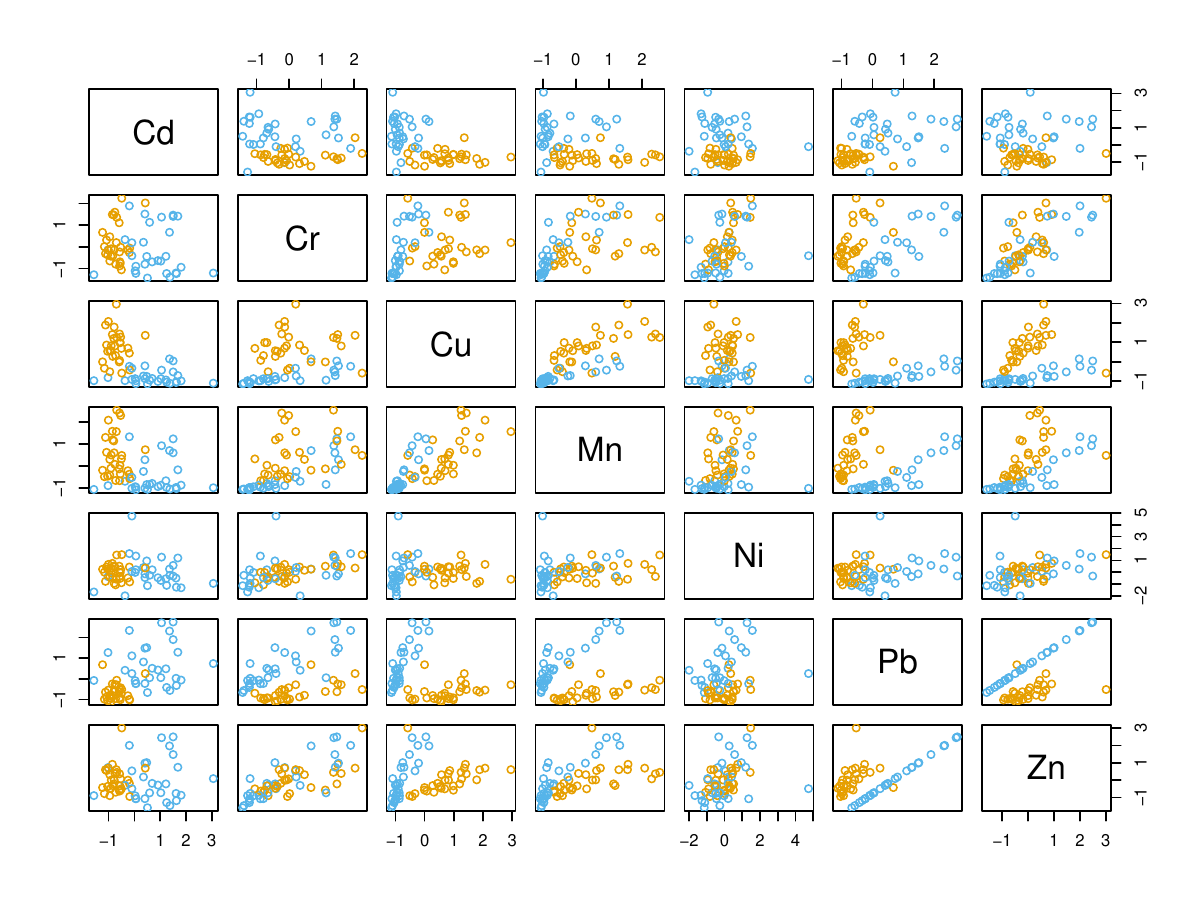}
    \caption{Scatter Plot}
    \label{fig:seaweed_scatter}
\end{subfigure}
\hfill
\begin{subfigure}[b]{0.4\textwidth}
    \centering
    \includegraphics[width=\textwidth]{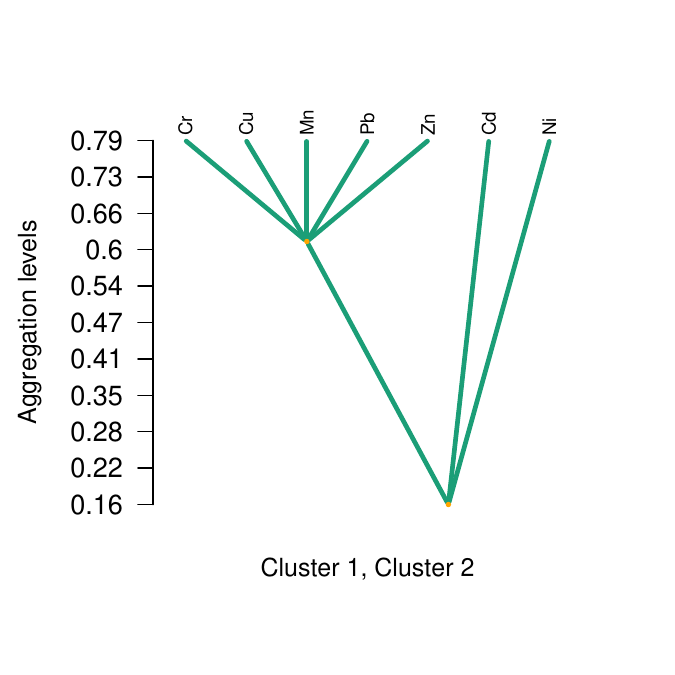}
    \caption{Path Diagram}
    \label{fig:seaweed_path}
\end{subfigure}
\caption{Pairs plot of the data where the observations are coloured by seaweed species (a) and path diagram corresponding to the PUMMM results (b).}
\label{fig:seaweed_combined}
\end{figure}

The Harbour Metals dataset \citep{roberts08} contains seven chemical concentration measurements for 60 seaweed observations collected from Sydney Harbour. Each observation belongs to one of two seaweed species: \textit{Padina crassa} (1) or \textit{Sargassum linearifolium} (2). Figure \ref{fig:seaweed_scatter} provides a visualization of the data. We fit the data for $G = 1, \hdots , 4$ for all methods, and $m  = 1, \hdots , 7$ for the PUMMMs and PUGMMs. The PUMMMs use the two-step model selection method introduced herein, while the PUGMMs and the Manly mixture models use the BIC, their default model selection criterion.

In Table~\ref{table:harbour_metals}, the proposed PUMMMs are the only method to select the theoretical number of clusters and achieve a competitive ARI.  Using the two-step model selection procedure, the PUMMMs select $G = 2$, $m = 2$, and the EUEE covariance case, grouping the seven chemicals into the two latent variable groups. This results in 11 covariance parameters, a significant reduction compared to 56 parameters in an unconstrained covariance matrix. The resulting ARI is 0.93, corresponding to a single misclassified observation. 

Looking at Figure~\ref{fig:seaweed_path}, chromium (Cr), copper (Cu), manganese (Mn), lead (Pb), and zinc (Zn) are in one group, with cadmium (Cd) and nickel (Ni) in the other. The former group exhibits a large within-group covariance of 0.61, reflecting strong internal concordance amongst these chemical elements. In contrast, the latter group displays a low within-group covariance of 0.16, equal to the between-group covariance. This indicates that cadmium and nickel are no more concordant with each other than with the chemicals in the other group. 

This result is consistent with \cite{roberts08}, who found that copper, manganese, zinc, and lead have the highest concentrations across both seaweed species, whilst cadmium, chromium, and nickel tended to occur at lower concentrations. Accordingly, the first latent group may be interpreted as representing the higher-concentration chemical elements, and the second as the lower-concentration chemical elements. A notable difference from \cite{roberts08} is that chromium is assigned to the higher-concentration group rather than the lower-concentration group, suggesting a deviation in how chromium associates with other chemical elements under this ultrametric framework, which may be of interest to domain experts.

\begin{table*}[!ht]
	\centering
 \small
	\caption{A comparison of the selected methods on the Harbour Metals dataset including $G$, $m$, case, number of free covariance parameters, $\rho_{\text{cov}}$, and ARI.}	
		\begin{tabular*}{1.0\textwidth}{@{\extracolsep{\fill}}lcccccc}
\toprule
Method & $G$ & $m$ & Case & $\rho_{\text{cov}}$ & ARI \\
\midrule
PUMMMs (Two-step) & 2 & 2 & EUEE & 11 & 0.93 \\
PUGMMs (BIC) & 3(2) & 4(6) & FFFI(EEEE) & 71(39) & 0.46(0.08)\\
Manly Mixture Model (Full) & --(2) & NA & NA & --(56) & --(0.58)\\
Manly Mixture Model (Forward) & --(--) & NA & NA & --(--) & --(--)\\
Manly Mixture Model (Backward) & --(--) & NA & NA & --(--) & --(--)\\
\bottomrule
	\end{tabular*}
	\label{table:harbour_metals}
\end{table*}

\section{Discussion}
\label{Sec6}

The PUMMM family, along with a two-step model selection procedure, has been introduced. These models simultaneously accommodate skewness, identify hierarchical relationships among groups of variables, and reduce the number of free covariance parameters relative to a Manly mixture model. The proposed models and model selection method are available in the {\sf R} package \texttt{PUGMM}.

Simulation studies and real data analyses demonstrate the effectiveness of both the PUMMMs and the proposed model selection procedure. The first simulation study highlights the strong performance of the two-step method, particularly in accurately detecting the number of latent groups of variables $m$. The second and third simulations demonstrate the strong clustering performance of the PUMMMs on Gaussian, Manly, and skew-$t$ simulated data. For several benchmark datasets, the PUMMMs achieved competitive clustering results with large ARIs and significant reductions in the number of free covariance parameters. A more detailed analysis of one of the benchmark datasets, Body, and of the Harbour Metals dataset demonstrates the ability of the PUMMMs to uncover a meaningful hierarchical structure amongst variables for each cluster.

For high-dimensional data, such as the Body and Wine~27 datasets, parallel implementation, an option in the $\texttt{PUGMM}$ package, is recommended, as estimation of the transformation vectors $\boldsymbol{\lambda}_g$ becomes computationally demanding as $p$ increases. Overall, the PUMMM framework provides a flexible, parsimonious, and interpretable approach to clustering skewed and high-dimensional data. 

\bibliographystyle{chicago}
\bibliography{ref.bib}

\appendix
\section{Extension of Simulation 1}

\label{app:sim1_ext}
In this extension, the two-step model selection method is implemented in the PUGMMs to explore how it performs in the Gaussian setting. As in Simulation~1, a total of 160 samples for each scenario are generated from a three-component Gaussian mixture model for $n = 250$ and 500. The four scenarios and parameter settings are identical to those described in Simulation 1. 

Table~\ref{table:sim1_ext} shows that the two-step model selection method consistently outperforms the BIC. Specifically, there is a significant increase in the percentage of times that the correct $m$, covariance case, and model ($G$, $m$, case) are selected using the proposed two-step approach. Looking at the results corresponding to $n = 500$, the percentage for the model ($G$, $m$, case) is low due to the overestimation of $G$. Overall, we see that the two-step model selection procedure selects the correct model more often than the BIC.
\begin{table*}[!ht]
	\centering
 \small
	\caption{A comparison of model selection approaches for PUGMM: the proposed two-step method (left) and the BIC (right). The table summarizes the amount of times that the correct $G$, $m$, case, and complete model ($G$, $m$, case) are selected.}	
	\scalebox{1}{
		\begin{tabular*}{1.05\textwidth}{@{\extracolsep{\fill}}cccccccccccc}
		  \noalign{\smallskip}\toprule
 		&&\multicolumn{2}{c}{$G$} &\multicolumn{2}{c}{$m$} &\multicolumn{2}{c}{Case} & \multicolumn{2}{c}{($G$, $m$, case)} \\ 
\cline{3-4}\cline{5-6}\cline{7-8}\cline{9-10}\cline{11-12}
		              Model& $n$ &Two-step&BIC&Two-step&BIC&Two-step&BIC&Two-step&BIC\\
\midrule
\multirow{2}{*}{EUUU}&$250$& 95.63\% & 95.63\% & 73.75\% & 45.63\% & 72.50\% & 45.63\% & 68.13\% &  43.75\% \\
&$500$ & 4.38\% & 5.63\% & 90.00\% & 47.50\% & 87.50\% & 45.63\% & 3.13\% & 3.13\% \\
\midrule
\multirow{2}{*}{EEEE}&$250$ & 96.88\% & 96.88\% & 93.75\% & 76.88\% & 94.38\% & 85.63\% & 86.88\% &  73.75\% \\
&$500$& 5.63\% & 5.63\% & 96.88\% & 88.75\% & 100.00\% & 100.00\% & 5.63\% & 5.00\% \\
\midrule
\multirow{2}{*}{FIII}&$250$ & 93.75\% & 93.75\% & 96.88\% & 79.38\% & 82.50\% & 75.63\% & 76.25\% &  70.00\% \\
&$500$& 5.00\% & 5.00\% & 98.75\% & 87.50\% & 91.88\% & 87.50\% & 4.38\% & 3.75\% \\
\midrule
\multirow{2}{*}{FFFF}&$250$& 93.75\% & 93.75\% & 88.75\% & 32.50\% & 25.63\% & 4.38\% & 20.00\% &  4.38\% \\
&$500$ & 6.25\% & 6.25\% & 95.63\% & 61.25\% & 27.50\% & 16.25\% & 3.75\% &  2.50\% \\
\bottomrule
	\end{tabular*}}
	\label{table:sim1_ext}
\end{table*}

\section{Sources for Benchmark Datasets}
Table \ref{table:realdata_source} contains the source information for the {\sf R} packages that contain the benchmark datasets used in Section~\ref{Benchmark}.
\begin{table}[!ht]
\centering
\caption{The {\sf R} packages and corresponding references for the benchmark datasets used in Section~\ref{Benchmark}.}
\begin{tabular}{l l l} 
 \toprule
Package & Version & Reference \\[0.5ex] 
 \hline
 \texttt{ContaminatedMixt} & 1.1 & \citet{ContaminatedMixt} \\
 \texttt{gclus} & 1.3.3 & \citet{gclus} \\
 \texttt{pgmm} & 1.2.7 & \citet{pgmm}\\
\texttt{ManlyMix} & 0.1.7 & \citet{ManlyMix}\\
\texttt{mclust} & 6.1.1 & \citet{mclust}\\
 \bottomrule
\end{tabular}
\label{table:realdata_source}
\end{table}

\end{document}